\documentclass[apj,twocolumn,twocolappendix]{openjournal}

\usepackage[utf8]{inputenc}
\usepackage{textcomp}
\usepackage{url}
\usepackage{graphicx}
\usepackage{epstopdf}
\usepackage{comment}
\usepackage{mathrsfs}
\usepackage{multirow}
\usepackage{booktabs}
\usepackage[export]{adjustbox}
\usepackage{orcidlink}
\usepackage{amsmath} 
\usepackage{hyperref}

\newcommand{\orcidauthor}[3]{%
  \author{\href{http://orcid.org/#1}{#2 \openin1 Orcid-ID.png \ifeof1 \else \hskip2pt\includegraphics[width=9pt]{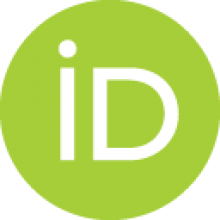}\fi}$^{#3}$}%
}

\usepackage{savesym}
\savesymbol{tablenum}
\usepackage{siunitx}

\restoresymbol{SIX}{tablenum}

\newcommand{\msun}{\rm M_{\odot}}

\hypersetup{
    colorlinks=true,
    linkcolor={red},
    citecolor={blue},
    urlcolor={blue},
    pdftitle={A new timing solution for PSR~J1807-2459B}
}

\begin{document}

\title{PSR~J1807$-$2459B in NGC 6544: A Unique Relativistic Binary in a Dense Cluster}

\orcidauthor{0000-0001-9929-2370}{A.~Carleo$^*$}{1}
\orcidauthor{0000-0001-5229-7430}{R.~Lynch}{2}
\orcidauthor{0000-0003-1307-9435}{P.\,C.\,C.~Freire}{3}
\orcidauthor{0000-0002-5924-3141}{A.~Corongiu}{1}
\orcidauthor{0000-0001-5799-9714}{S.\,M.~Ransom}{4}
\orcidauthor{0000-0003-2444-838X}{F.~Kareem}{3}
\orcidauthor{0000-0002-9791-7661}{F.~Abbate}{1}
\orcidauthor{0000-0002-1823-3571}{M.~Colom i Bernadich}{1,3}
\orcidauthor{0009-0000-2963-6641}{K.~Grunthal}{3}
\orcidauthor{0000-0001-9784-8670}{I.\,H.~Stairs}{5}
\orcidauthor{0000-0003-2317-1446}{J.\,W.\,T.~Hessels}{6,7,8,9}
\orcidauthor{0000-0002-2185-1790}{M.\,E.~Decesar}{10}
\orcidauthor{0000-0001-6762-2638}{A.~Ridolfi}{11}
\orcidauthor{0000-0001-5902-3731}{A.~Possenti}{1}
\orcidauthor{0009-0004-7250-6032}{J.~Debnath}{1}
\orcidauthor{0000-0002-1806-2483}{D.~Perrodin}{1}

\affiliation{$^1$ INAF -- Osservatorio Astronomico di Cagliari, Via della Scienza 5, I-09047 Selargius (CA), Italy \\ $^*$ e-mail: amodio.carleo@inaf.it }
\affiliation{$^2$ Green Bank Observatory, P.O. Box 2, Green Bank, WV 24944, USA}
\affiliation{$^3$ Max-Planck-Institut für Radioastronomie, Auf dem Hügel 69, D-53121 Bonn, Germany}
\affiliation{$^4$ National Radio Astronomy Observatory, 520 Edgemont Road, Charlottesville, VA 22903, USA}
\affiliation{$^5$ Department of Physics and Astronomy, University of British Columbia, 6224 Agricultural Road, Vancouver, BC V6T 1Z1 Canada}
\affiliation{$^6$ Department of Physics, McGill University, 3600 rue University, Montr\'eal, QC H3A 2T8, Canada}
\affiliation{$^7$ Trottier Space Institute, 3550 rue University, Montr´eal, QC H3A 2A7, Canada}
\affiliation{$^8$ Anton Pannekoek Institute for Astronomy, University of Amsterdam, Science Park 904, 1098 XH Amsterdam, The Netherlands}
\affiliation{$^9$ ASTRON, Netherlands Institute for Radio Astronomy, Oude Hoogeveensedijk 4, 7991 PD Dwingeloo, The Netherlands}
\affiliation{$^{10}$ Department of Physics and Astronomy, George Mason University, Fairfax, VA 22030, USA}
\affiliation{$^{11}$ Fakultät für Physik, Universität Bielefeld, Bielefeld 33501, Germany}

\begin{abstract}
The globular cluster NGC\;6544 is a high-density, dynamically evolved system located in the direction of the Galactic Center, known for hosting the eccentric binary millisecond pulsar PSR~J1807$-$2459B. We present a new timing solution for this pulsar, based on more than 26 years of observations with the Parkes, Green Bank and MeerKAT radio telescopes. We measure for the first time the variation of the orbital period, which is dominated by the acceleration of the binary in the cluster potential: this places the pulsar in front of the cluster center and reveals its intrinsic spin-down, typical of a fully recycled millisecond pulsar. We also improve the measurements of the rate of advance of periastron and of the Einstein delay, the latter by a factor of four, and include the previously neglected contributions of the proper motion, which makes these parameters and the derived masses of the pulsar and its companion, $1.3636^{+0.0017}_{-0.0021}$ and $1.2071^{+0.0021}_{-0.0017}\,{\rm M_\odot}$, more realistic. Together with the Shapiro delay, these measurements allow two independent tests of general relativity, both passed. Our results show that precise post-Keplerian parameters of a binary pulsar in a dense cluster can probe both the cluster gravitational field and general relativity, provided that kinematic effects are carefully taken into account.
\end{abstract}


\maketitle

\section{Introduction}


Globular clusters (GC) are ancient, very dense star systems that are gravitationally bound to galaxies and contain roughly $10^4$ to $10^6$ stars \citep{Gratton_2019}. Because they are old and dynamically crowded, they preserve important clues about the early formation history of galaxies and are excellent laboratories for stellar evolution. Although they were once thought to be simple stellar populations, high-precision observations now show that nearly all globular clusters contain multiple stellar populations with different light-element abundances, indicating a more complex chemical history \citep{Bastian_2018}. Their high central densities also drive frequent close encounters, exchanges, and collisions, which produce X-ray binaries and especially millisecond pulsars (MSP). If the interaction rate per binary \citep{Verbunt_2014} is large, then multiple exchange encounters for the same neutron star become possible. This can lead to the formation of exotic systems, such as possible MSPs with massive companions in eccentric orbits - such as the target of this study. This class can even contain pulsar-black hole systems \citep{Barr2024}.
 
Studying pulsars in globular clusters therefore provides important information on binary evolution, neutron-star recycling, cluster dynamics, on the retention of compact remnants and provides valuable laboratories for tests of General Relativity (GR) and alternative gravity theories \citep{Freire_2024_tests_gr}. However, this can be limited by intrinsic difficulties in the timing. More specifically, the line‐of‐sight acceleration  due to the cluster’s gravitational potential can alter the observed spin period derivative  as well as the orbital period derivative. Inferring intrinsic spin-down (and hence characteristic age, magnetic field) thus requires detailed modeling of cluster mass distribution and knowledge of the pulsar’s location within the cluster \citep{prager_2017,Freire_2017_F23}. Moreover, in very dense cluster cores, close passages of other stars or compact objects may perturb the orbit (changing orbital period, eccentricity) or cause timing noise (irregularities in pulse time of arrival due to gravitational perturbation or varying propagation delays). A further challenge stems from the variations in dispersion measure (DM), interstellar scattering, and possibly intra-cluster medium ionized gas which can introduce delays. For some pulsars, pulse broadening or scattering might smear pulses or introduce frequency-dependent delays, complicating multi-frequency timing. Last but not least, the detection probability can be very low, especially for clusters in crowded fields near the Galactic plane, where dispersion and scattering in the interstellar medium can severely affect the signal. Additional detectability losses may arise from intrinsic binary eclipses or absorption. \\


The cluster NGC\;6544 is a particularly compelling system to study. As reported in the Globular Cluster database\footnote{\url{https://people.smp.uq.edu.au/HolgerBaumgardt/globular/}}, it is located at a heliocentric distance of $2580\pm60$pc, according to  N-body model fits to Gaia proper motions, HST photometry, and radial velocity data \citep{Baumgardt_2018, Vasiliev_2019}.
From the multi-band photometric study of \cite{Cohen_2014}, the heliocentric distance  of NGC\;6544 was derived as $\sim2.47$~kpc, but significantly lower than assumed in the previous work on PSR~J1807$-$2459B \citep{lynch2012}, i.e. 3 kpc.

The total mass is estimated to be  $(8.15 \pm 0.56) \times10^4 \, \msun$ and the mass-to-light ratio is $M/L_V\approx2.30\pm0.52 \, \msun/L_\odot$  \citep{Baumgardt_2018}. The database also reports an half-mass radius of $r_h\approx2.24\,$pc, with a total stellar mass function slope of $\alpha\approx -0.49\pm0.19$ (for stars with  mass in the range $0.38 \, \msun$ to $0.82 \, \msun$), which is much flatter than a canonical initial mass function slope of about $-1.3$ to $-2.3$. This means that NGC 6544 is depleted in low-mass stars relative to high-mass ones (bottom-light mass function). Such flattening typically results from dynamical evolution - preferential escape of low-mass stars due to tidal stripping and two-body relaxation.  As a result, the cluster  is strongly mass-segregated. This is confirmed by the fact that  it is relatively very compact and has an estimated age of 11-13 Gyr \citep{2017A&A...608A.140C,Baumgardt_2023}, which places it among the oldest known objects in the Milky Way galaxy.   It also experiences notable reddening due to its position — being close to the Galactic plane/bulge — complicating optical observations.
This globular cluster also has one of the highest interaction rates per binary of any GC, which indicates a heightened probability of finding exotic binaries. 
 
NGC\;6544 hosts  three millisecond binary radio pulsars, namely PSR J1807$-$2459A \citep{Damico_2001, Ransom_2001, lynch2012}, PSR J1807$-$2459B \citep{lynch2012} and PSR J1807$-$2459C\footnote{\url{https://www.trapum.org/discoveries/}} (see Tab.\,\ref{tab:1}). PSR J1807$-$2459A, is a $\sim3.06$ ms pulsar in a very tight, nearly circular orbit (orbital period $P_b\approx0.0711$ d) with a very low‐mass companion (minimum mass $m_2\gtrsim0.009 \, \msun$), characteristic of a “black widow” system \citep{Ransom_2001, Damico_2001,  lynch2012}.
PSR J1807$-$2459B, the target of this study, is a $\sim4.186$ ms pulsar in a large ($P_b\approx10$ d) and very eccentric orbit ($e \simeq 0.75$), with a massive companion ($m_2 \simeq 1.2\, \msun$). Its spin, orbital and mass configuration strongly suggest this is a "secondary"\footnote{The word "secondary" is used to distinguish this encounter from the "primary" exchange that formed the low-mass X-ray binary in which the MSP was formed.} exchange system, where the companion is not the remnant of the star that recycled the pulsar. This is the type of system that is expected to occur only in the GCs with the highest interaction rates per binary. Its identification as a secondary exchange product is based on the fact that MSPs occur in circular orbits with $m_2<0.5\,\msun$ companions, being formed from long-lasting recycling from a low-mass, slowly evolving companion during a low-mass X-ray binary stage \citep{2023pbse.book.....T}. There are similarly eccentric binary pulsars with massive companions in the Galaxy, but these are generally double neutron star systems. However, owing to the larger companion masses, the progenitors of such companions evolve much faster. This implies that neutron stars in such systems undergo much shorter recycling episodes, resulting in pulsars with an order of magnitude longer spin periods \citep{tauris2017dns}. Finally, PSR~J1807$-$2459C has been discovered more recently in surveys by MeerTIME/TRAPUM, with a period of 5.73 ms \citep{Meertime_2016, Trapum_2016}.  There are no known isolated pulsars in NGC\;6544 published in the literature as of now. \\


In this work, we present a new timing solution for PSR~J1807$-$2459B, based on more than 26 years of observations, which extends and improves that of \cite{lynch2012}. The paper is organized as follows: Sec.~\ref{sec:data} describes the observations and data processing; Sec.~\ref{sec:timing} presents the timing analysis, including the noise modeling; Sec.~\ref{sec:discussion} discusses the orbital period derivative and its implications for the pulsar and the cluster, the other post-Keplerian parameters, the masses of the system and the tests of GR; finally Sec.~\ref{sec:summary} summarizes our results. \\

\begin{table*}[ht]
\caption[]{\label{tab:1} Pulsars in NGC\;6544 and their main parameters. From the left to the right, offset from the cluster center, spin period, spin period derivative, dispersion measure, orbital period, projected semi-major axis, eccentricity and companion mass for the three (binary) pulsars discovered so far in NGC\;6544. Notice that for J1807$-$2459C a timing solution is still not available. For the references: (a) \cite{Damico_2001},  (b) \cite{Ransom_2001}, (c) \cite{lynch2012}, (d) \href{https://www.trapum.org/discoveries/}{TRAPUM}.}
\centering
\resizebox{\textwidth}{!}{%
\begin{tabular}{l|cccccccccl}
\hline
\hline
Name &
Offset &
Period &
$\dot{P}$ &
DM &
Orbital Period &
$x$ &
Ecc &
$m_2$ &
References \\
 & (arcmin) & (ms) & ($10^{-20}$) & (cm$^{-3}$\,pc) & (days) & (lt-s) &  & ($\msun$) &  &  \\
\hline
J1807$-$2459A & 0.07 & 3.06 & -0.43 & 134 & 0.07 & 0.01 & 0.0 & 0.01 & a, b, c \\
J1807$-$2459B & 0.20 & 4.19 & 8.24 & 137 & 9.96 & 28.92 & 0.75 & 1.2  & c, this work \\
J1807$-$2459C & * & 5.73 & * & 136 & * & * & * & *  & d\\
\hline
\hline
\end{tabular}}
\end{table*}

\section{Observations and data processing}\label{sec:data}

\subsection{Observations}

We used search-mode data collected by  Parkes, Green Bank Telescope (GBT),  
and MeerKAT, spanning from 1999 to 2026.
A summary of all observations we used in this work is provided in Tab.~\ref{tab:2}. In particular, we used  Parkes data from 1999 to 2012 recorded with the Analogue Filter Bank (AFB) at a central frequency of 1390 MHz, as well as Parkes data from 2013 to 2016 recorded with the Digital Filter Bank 4 (DFB4) at a similar central frequency. Both datasets have a limited bandwidth of 256 MHz and used the 13-beam 21 cm Multibeam receiver \citep{Multibeam}. The AFB divided the band into 512 channels of
0.5\,MHz each, while the DFB4 data have 1024 frequency
channels. The AFB data have a data span of 12.82 years with a mean cadence of $\sim$48 days and a median cadence of $\sim$29 days, with a maximum interval of 1.16 years, from 31 July 2006 to 29 September 2007. The seven DFB4 observations span 2.4 years, with a mean interval between consecutive observations of $\sim$145 days (median $\sim$141 days).

GBT observations span from 2004 to 2026. The earliest data consist of three
archival observations taken in July 2004 with the prime-focus PF1 receiver at
825\,MHz, recorded with the Berkeley-Caltech Pulsar Machine (BCPM2)
\citep{BCPM} with a bandwidth of 48\,MHz divided into 96 channels.  
All subsequent GBT observations were made with the L-band and S-band receivers,
which have nominal frequency ranges of 1.15--1.73\,GHz and 1.73--2.6\,GHz,
respectively. From 2009 to 2020 the data were recorded with the Green Bank
Ultimate Pulsar Processing Instrument (GUPPI), after which we began using the
Versatile Green Bank Astronomical Spectrometer (VEGAS) backend; in both cases,
data were recorded in PSRFITS search mode. The GUPPI observations taken at
S band between February 2009 and January 2010, and at L band between February
and July 2011, are those presented by \cite{lynch2012}. 
The 2009--2010 data were recorded with incoherent dedispersion and 2048
frequency channels across a bandwidth of 800\,MHz, while all observations from
2011 onwards were recorded with coherent dedispersion. 
From 2011, all GUPPI data and a subset of the VEGAS data had a sampled bandwidth
of 800\,MHz with 512 frequency channels. In 2022 we began sampling a bandwidth
of 1500\,MHz using 1024 frequency channels when using the S-band receiver
(L-band observations continued to sample 800\,MHz bandwidth). Note, though, that
the effective bandwidth is typically limited by the receiver to a value lower
than the sampled bandwidth. Since late 2011, the observing cadence, considering
L- and S-band observations together, varied but was typically about 90 days,
and observation durations were typically about two hours\footnote{The earlier
campaigns of \cite{lynch2012} were considerably denser.}. 

Finally, we used MeerKAT search-mode data from 2021 to 2023 performed with two receivers: two observations with the L-band receiver which covers the frequency range 0.86-1.71 GHz (one with 512 channels and one with 256 channels) and one observation with the
S1-band receiver which covers the frequency range 1.97-2.84 GHz (with 256 channels).  The first two were performed  in February 2021 and September 2022 respectively, while the S1-band observation was recorded in July 2023. 

In total, we have collected approximately 103 hours of observations from Parkes, 184 hours from GBT, and 10 hours from MeerKAT. Combined, the observations span an extensive time period, with a data baseline of about 26 yr, without significant temporal gaps.  Concerning the orbital phase (computed from the periastron), Parkes and the majority of the GBT observations were scheduled without considering it, and therefore they show a well-distributed coverage. On the other hand, the limited MeerKAT observations cover very restricted orbital portions: the two L-band observations  are clustered in the ranges 0.19-0.21 and 0.97-0.99 respectively, whereas the S1-band observation spans the narrow orbital-phase interval  $\sim$0.84-0.85.

Comparing the different datasets, the MeerKAT and GBT observations generally show lower noise and provide higher-quality profile measurements than the Parkes data. This is reflected in their smaller ToA (Time of Arrival) uncertainties, typically a factor of a few lower than for Parkes. Nevertheless, the Parkes observations provide the earliest measurements and allow us to compare the pulse-profile structure at different epochs and at frequencies overlapping the MeerKAT L-band receiver. The long temporal coverage provided by Parkes, combined with the extensive and relatively precise GBT dataset, is particularly important for measuring secular variations of the Keplerian parameters. The broad frequency coverage of both GBT and MeerKAT also provides strong constraints on the  DM evolution. Moreover, unlike the limited MeerKAT observations, the GBT data provide a much broader orbital-phase coverage, including measurements closer to superior conjunction, where the Shapiro-delay signature is strongest.


\begin{table*}
\caption{\label{tab:2}
Summary of the datasets used in this analysis for PSR J1807$-$2459B.
The data were recorded in search mode with one polarization channel.
The system temperature values are drawn from the
\href{https://www.atnf.csiro.au/research/multibeam/lstavele/description.html}
{Multibeam Receiver Description webpage},
the \href{https://greenbankobservatory.org/wp-content/uploads/2017/06/GBTog.pdf}{Green Bank Telescope receiver documentation}, and the
\href{https://skaafrica.atlassian.net/wiki/spaces/ESDKB/pages/277315585/MeerKAT+specifications}
{MeerKAT General documentation}.
The gain is estimated from the collecting area, accounting for only
56 antennas being available in the MeerKAT S-band configuration.
The last two columns indicate timing-data reduction parameters:
the number of ToAs per dataset and the median ToA uncertainty. 
Notice that, throughout  the manuscript, the GUPPI datasets are identified by different timing flags: observations from 14 Feb.\ 2009 to 4 Jan.\ 2010 are labelled \texttt{INCO-GUPPI\_SBAND}, those from 15 Feb.\ to 18 Jul.\ 2011 as \texttt{CO-GUPPI\_LBAND}, while all subsequent GUPPI observations, from 29 Sep.\ 2011 to 29 Mar.\ 2020, are labelled \texttt{GUPPI\_L} (L-band) and \texttt{GUPPI\_S} (S-band). 
}
\centering
\resizebox{\textwidth}{!}{%
\begin{tabular}{l|ccccccccccc}
\hline
\hline \\[-1.5ex]
Receiver(s) & Backend & Freq. & Bandw. & Temp. & Gain &
\# obs & Time & First obs. & Last obs. & \# ToAs & Median err. \\[0.3ex]
& & (MHz) & (MHz) & K & K\,Jy$^{-1}$ &
& (hours) & & & & ($\mu$s) \\
\hline \\[-1.5ex]

Parkes Multibeam
& AFB
& 1390
& 256
& 21
& 0.9
& 98
& 91.8
& 9 Dec. 1999
& 5 Oct. 2012
& 200
& 25.15 \\[0.5ex]

Parkes Multibeam
& DFB4
& 1369
& 256
& 21
& 0.9
& 7
& 11.5
& 20 Sep. 2013
& 7 Feb. 2016
& 15
& 15.46 \\[0.5ex]

GBT PF1 & BCPM2
 & 825 & 48 & 29 & 2.0
 & 3 & 3.5
 & 21 Jul. 2004 & 23 Jul. 2004
 & 14 & 51.54 \\[0.5ex]

GBT S-band
& GUPPI
& 2000
& 800
& 22
& 2.0
& 15
& 12.5
& 14 Feb. 2009
& 4 Jan. 2010
& 78
& 7.61 \\[0.5ex]

GBT L-band
& GUPPI 
& 1500
& 800
& 20
& 2.0
& 13
& 40
& 15 Feb. 2011
& 18 Jul. 2011
& 194
& 5.63 \\[0.5ex]

GBT L-band
& GUPPI
& 1500
& 800
& 20
& 2.0
& 22
& 44
& 29 Sep. 2011
& 30 Sep. 2019
& 794
& 9.24 \\[0.5ex]

GBT S-band
& GUPPI
& 2000
& 800
& 22
& 2.0
& 22
& 39
& 15 Apr. 2013
& 29 Mar. 2020
& 698
& 9.36 \\[0.5ex]

GBT L-band
& VEGAS
& 1500
& 800
& 20
& 2.0
& 8
& 20
& 23 Sep. 2020
& 13 Apr. 2026
& 367
& 11.28 \\[0.5ex]

GBT S-band
& VEGAS
& 2000
& 800/1500
& 22
& 2.0
& 11
& 25
& 29 Jun. 2020
& 10 Jan. 2026
& 521
& 10.09 \\[0.5ex]

MeerKAT L-band
& PTUSE
& 1284
& 856
& 18
& 2.6
& 2
& 8
& 7 Feb. 2021
& 19 Sep. 2022
& 140
& 8.29 \\[0.5ex]

MeerKAT S1-band
& PTUSE
& 2406
& 875
& 22
& 2.3
& 1
& 2
& 1 Jul. 2023
& 1 Jul. 2023
& 72
& 7.48 \\[0.5ex]

\hline
\end{tabular}}
\end{table*}


\subsection{Data processing}

Following standard data reduction procedures in pulsar timing, we used the \texttt{PSRCHIVE} \citep{2004_hotan} software package. If not explicitly indicated otherwise, all programs or commands referred to in this section are part of this package. 

All observations used in this work were in search-mode, therefore we first folded them using the command \texttt{dspsr} to perform offline folding of pulsar data into sub-integrations and frequency channels using some  ephemeris. Apart from Parkes-AFB data, which we folded with ephemeris taken by \cite{lynch2012}, to fold a specific dataset we used as ephemeris a  timing solution obtained with \texttt{TEMPO2} \citep{Hobbs:2006zg} using the previous datasets.  After obtaining a first, global timing solution, which phase-connected the ToAs across the complete timing baseline, we updated the ephemeris in all available observations. With the new ephemeris installed, we re-folded all datasets  to obtain better profiles and standard templates and then we  re-calculated the ToAs, as described in Sec.\ 3.1.   

We folded the AFB dataset setting 20-seconds  sub-integrations and 128 pulse phase bins. The number of frequency channels was 512, corresponding to the native number of frequency channels. The archives were manually excised of radio-frequency interference (RFI) with the \texttt{pazi} interface, and 
some bottom channels and  top channels were zero-weighted with the \texttt{paz} command to remove Gaussian noise caused by the loss of sensitivity at the edge of the bandpass. To optimize the DM value, we ran \texttt{pdmp} on each archive, obtaining an estimate which maximized the signal-to-noise  ratio (SNR). 


A very similar approach has been used for the remaining datasets. However,  some of the GBT data were folded using \texttt{fold\_psrfits} (part of the \texttt{psrfits\_utils} package\footnote{\url{https://github.com/demorest/psrfits_utils}}), while more recent data were folded with \texttt{dspsr}.  For these data, we used 256 pulse phase bins and 200-second sub-integrations. Finally, we folded  the MeerKAT data with 10-second sub-integrations. \\


\subsection{Profile frequency evolution} 

\begin{figure*}[htbp]
  \centering
  
  \begin{minipage}[t]{0.48\textwidth}
    \centering
    \includegraphics[width=\textwidth,trim={79 90 90 78},clip]{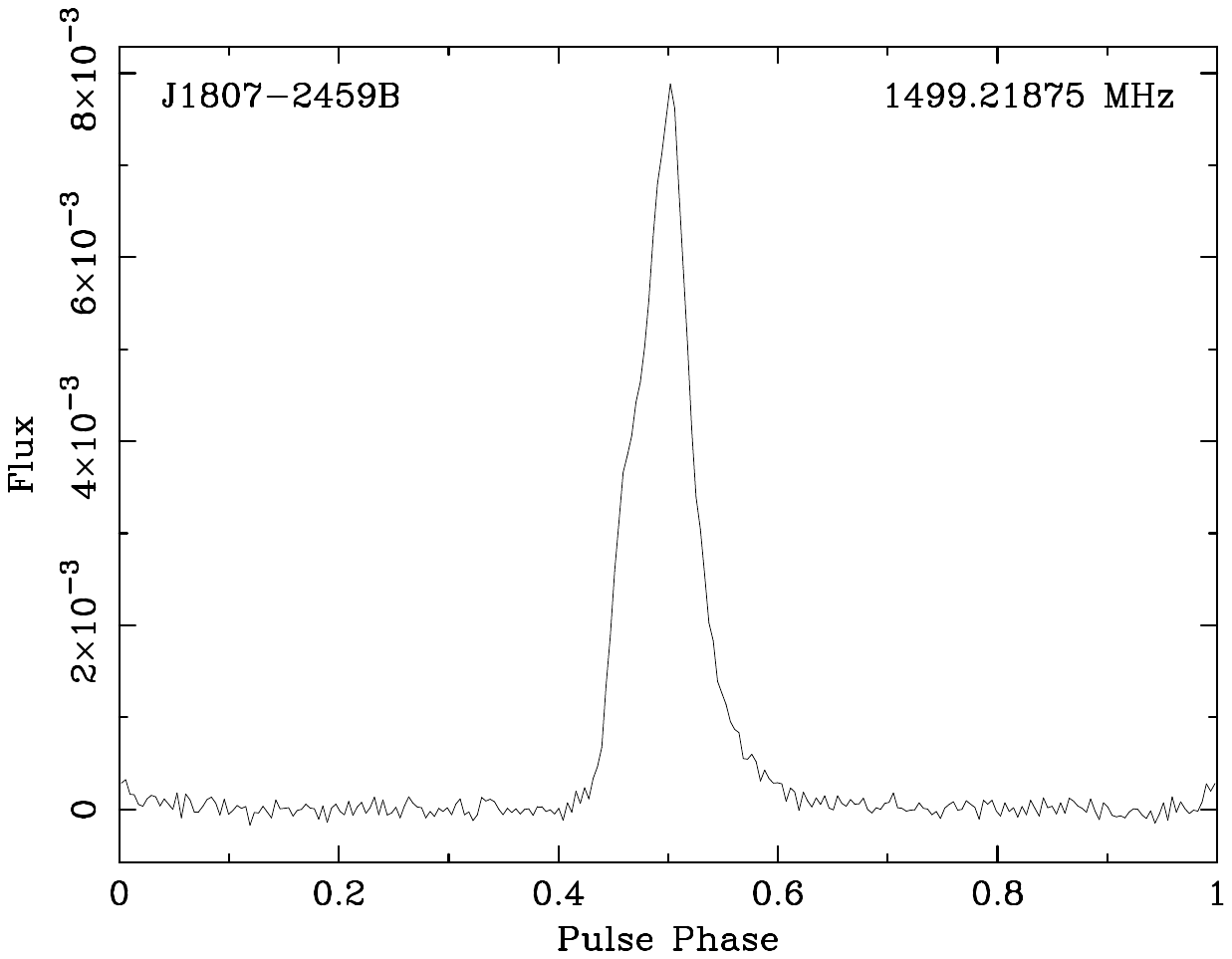}\par\nointerlineskip
    \includegraphics[width=\textwidth,trim={78 40 90 94},clip]{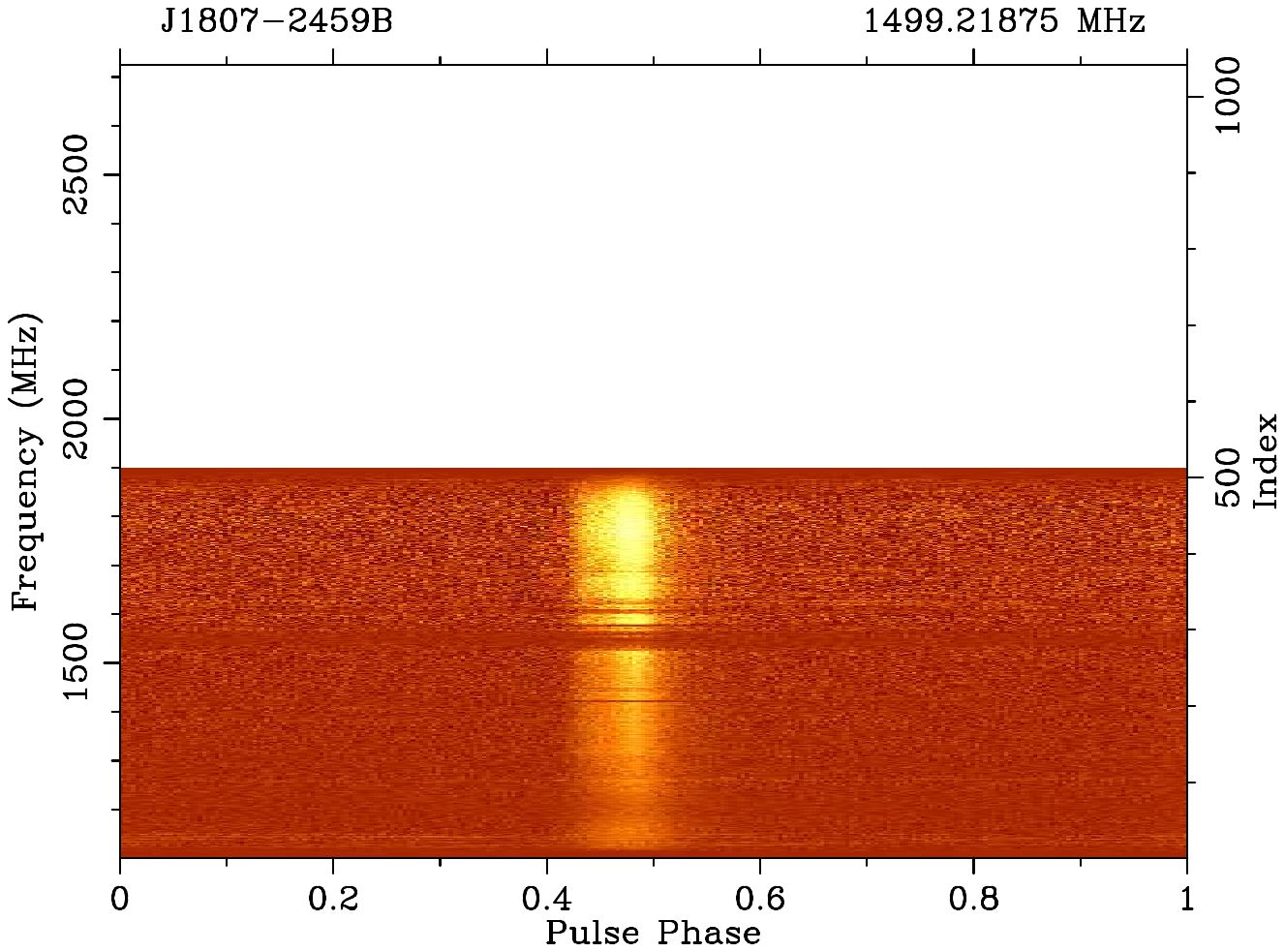}\par
    \vspace{2mm}
    (a)
  \end{minipage}\hfill
  \begin{minipage}[t]{0.48\textwidth}
    \centering
    \includegraphics[width=\textwidth,trim={78 90 90 90},clip]{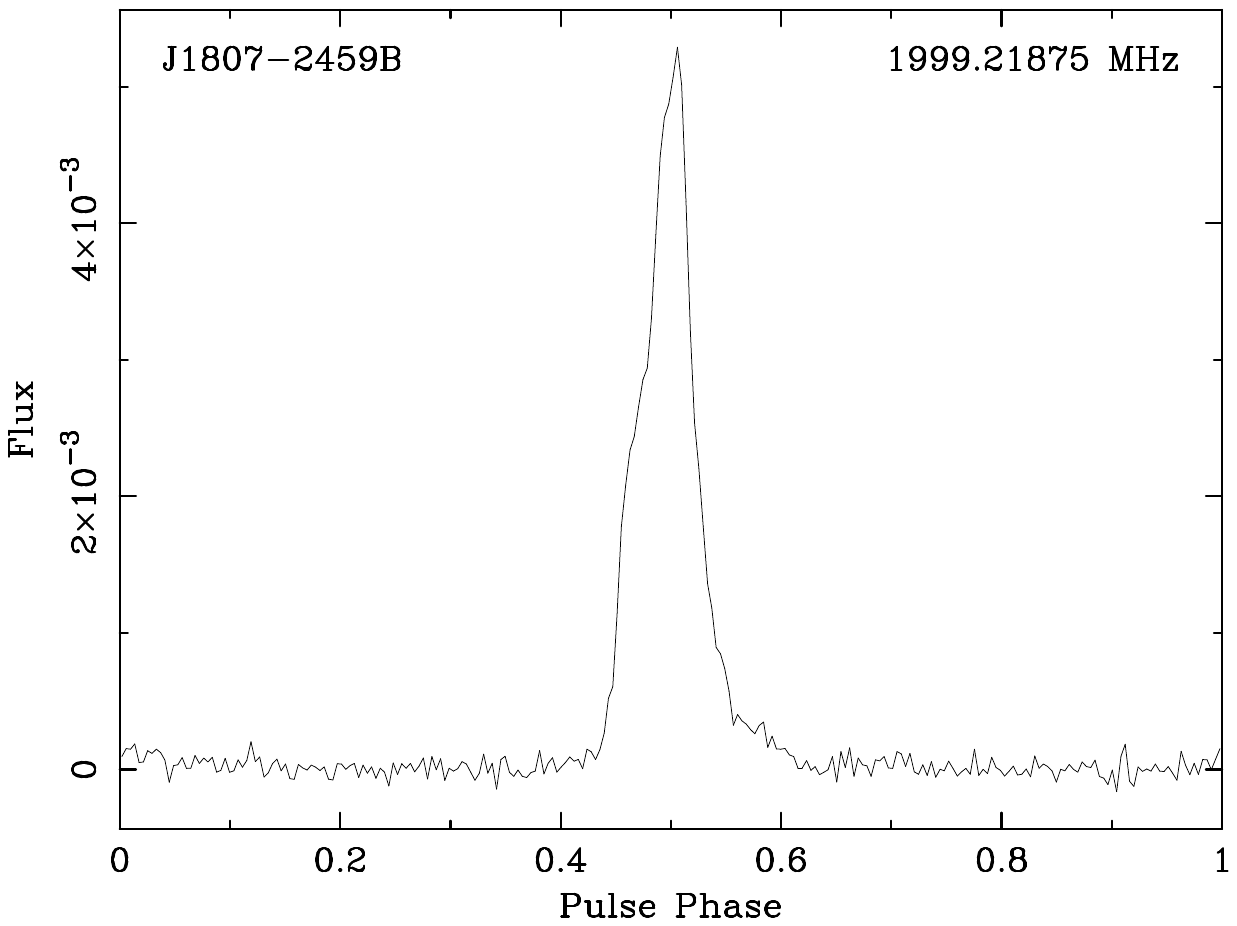}\par\nointerlineskip
    \includegraphics[width=\textwidth,trim={78 40 90 94},clip]{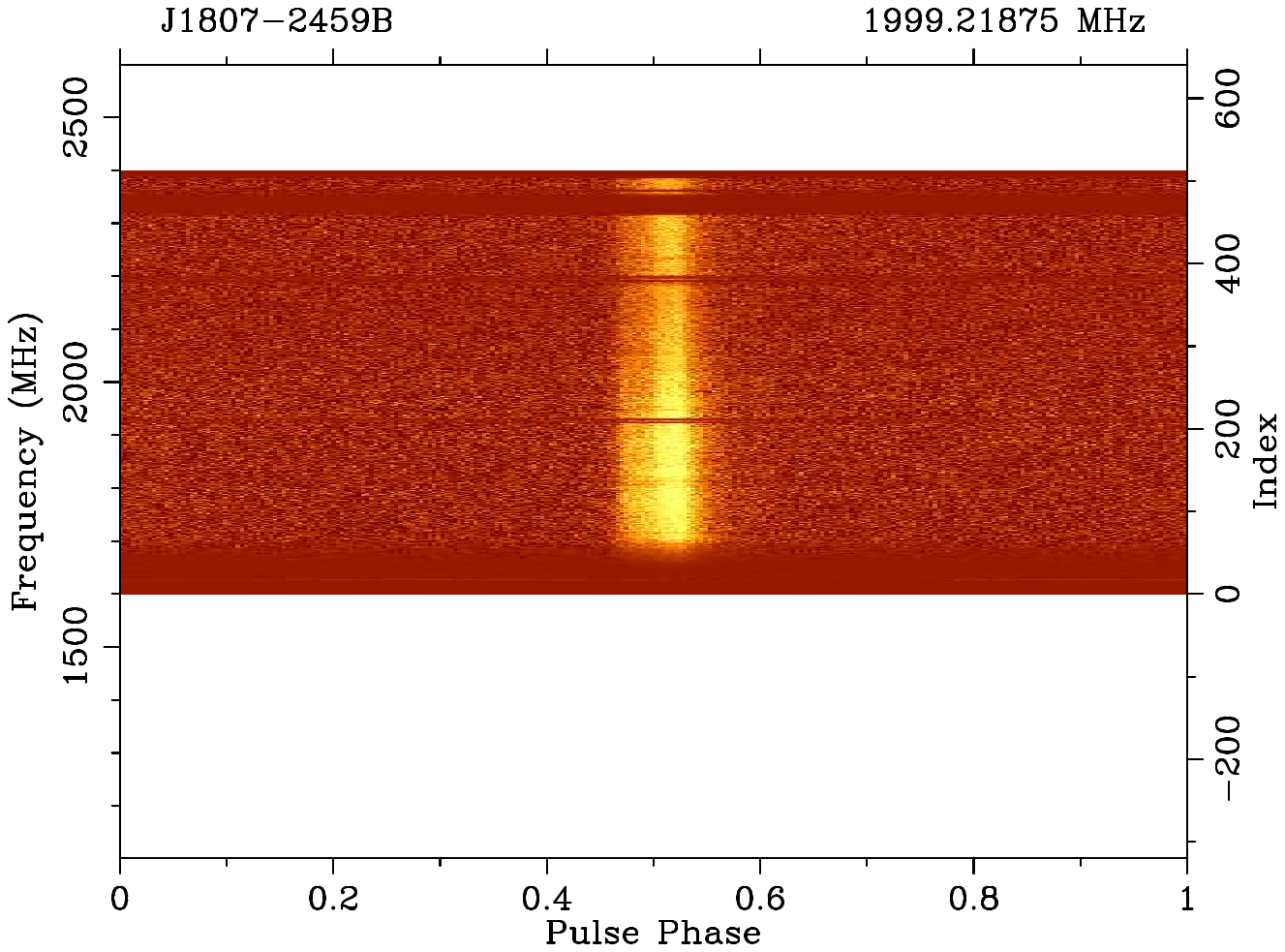}\par
    \vspace{2mm}
    (b)
  \end{minipage}

  \vspace{4mm} 

  \caption{Frequency resolved intensity profiles of PSR~J1807$-$2459B from the GBT/GUPPI
observations at (a) L band ($T_{\rm obs}\simeq44$\,h) and (b) S band
($T_{\rm obs}\simeq39$\,h), obtained by summing all the observations of each
dataset,  with brighter colors
indicating a higher received signal power. The top panel of each plot shows the total intensity profile across one period. Flux (density) is not calibrated and arbitrary units are assumed; therefore neither the vertical scales of the top panels nor the color scales of the bottom panels can be compared between (a) and (b). The center
frequency of each band is reported in the top-right corner.}
  \label{fig:1}
\end{figure*}

We used our global timing solution to obtain the pulse profile as a function of
frequency. For each dataset, the integrated profile was obtained by summing all
observations of J1807$-$2459B and integrating along the time and frequency axes
(no polarization information was available). Figure\;\ref{fig:1} shows the profiles obtained from the GUPPI L-band and S-band datasets. These data combine a high SNR with good time
resolution and wide frequency coverage. Since they were recorded with the same
backend, they also allow a direct comparison between the two bands. The bottom
panels show the profile as a function of frequency. Since
data are not flux-calibrated, the received signal power is in arbitrary units.

The pulse profile shows a single main pulse with a full width at half maximum
$W_{50}\simeq0.24$\,ms at L band, corresponding to a duty cycle of about
6 per cent; no interpulse is detected. The profile is very similar in the two
bands, the S-band one being only slightly narrower ($W_{50}\simeq0.21$\,ms).

In both bands, the pulse is clearly detected over the
whole usable bandwidth. Relative to the noise, we do not identify significant changes in the received
signal power across either band, although the pulse becomes slightly fainter
towards higher frequencies. 
The pulse appears dimmer at
the lower frequencies of the L band; however, since the noise level is correspondingly
lower there, this likely reflects the frequency response of the observing system.



The profiles from the other datasets are consistent with those in Fig.~\ref{fig:1} once their different SNR and time resolution are taken into
account. In particular, the MeerKAT S1-band observation, although limited to
2\,h, shows that the pulse is still detected up to $\sim$2.8\,GHz, with
no significant change of its shape across the band. Given the small profile evolution within each band, a single frequency-averaged
template was used for each dataset

\section{Timing Analysis}\label{sec:timing}

\subsection{Generating Times-of-Arrival}

We produced the ToAs for all datasets using the standard template-matching technique employed in pulsar timing. The ToAs were calculated by correlating a standard profile with the observed pulse profiles obtained from each observation archive, after integrating over polarization and over an appropriate range of time and frequency channels. 


For each dataset, the standard profile was built by summing with
\texttt{psradd} all observations with SNR above a minimum threshold and
scrunching the result in frequency and time with \texttt{pam}. The ToAs were obtained with \texttt{pat}, using its default algorithm, after scrunching each archive to a number of sub-bands and sub-integrations chosen according to its SNR.

The lower quality of the signal in the Parkes and GBT/BCPM2 observations resulted in larger ToA
uncertainties for those datasets. For these data, and especially for AFB, most ToAs were generated
using fully integrated observations, i.e. using one sub-band and sub-integration per observation. For the other observations, instead, we used 2 or 4 sub-bands for frequency-resolved ToAs. For the MeerKAT datasets, given the high SNR, we extracted 4 frequency-resolved and 18 time-resolved ToAs per observation, resulting in $\sim$70 ToAs per observation. To reduce to computational load of the timing analysis, we carefully discarded observations with a very low SNR  as well as a small number of ToAs with remarkably large uncertainties. Altogether, we obtained 3093 ToAs. \\

To analyze the ToA dataset, we used the timing software package \texttt{TEMPO2}, which implements a least-square minimization of the residuals based on the $\chi^2$  statistic. The different datasets were combined by introducing a jump between each of them, using the more recent VEGAS ToAs as a reference. These jumps were treated as
free fitting parameters in the \texttt{TEMPO2} fit. In total, we introduced 10 jumps to combine our datasets. 

For the binary orbit, \texttt{TEMPO2} provides several models based on the calculations by \cite{Damour1986_II} which provided a standard orbital model (henceforth "DD"). In this model, the orbital motion is parameterized by five Keplerian parameters (binary orbital period $P_\mathrm{b}$, longitude of periastron $\omega$, time of periastron passage $T_0$, orbital eccentricity $e$, and orbital semi-major axis projected along the line-of-sight $x$) and a number of additional post-Keplerian parameters that quantify, in a theory-independent way, the relativistic deviations from the Keplerian orbital motion. They are the rate of change of the orbital period $\dot{P}_\mathrm{b}$, the rate of periastron advance $\dot{\omega}$, the Einstein delay $\gamma$ (which quantifies the effects of the varying gravitational redshift and special-relativistic time dilation) as well as the Shapiro delay (which affects the propagation time of the radio waves to Earth). In the DD model, the latter effect is parameterized using the "range" $r$ and "shape" $s$ parameters . In GR, these are related to the companion mass $m_2$ and the sine of the orbital inclination angle $i$ respectively \citep{Damour1992_tests_of_gravity}. \\

To derive a timing solution for J1807$-$2459B we analyzed the ToAs with the theory-independent DDH model developed by \cite{DDH_model_2010}, which differs from the DD model only in the parameterization of the Shapiro delay, with the new post-Keplerian (PK) parameters ($h_3$ and $\varsigma$) which are less correlated than $r$ and $s$. However, we did not notice any significant difference in the measure of the Shapiro-related parameters between DD and DDH models, probably due to the high orbital inclination of J1807$-$2459B, a geometrical condition in which the correlation between  $r$ and $s$ is not strong.

\subsection{Noise analysis}
\label{sec:noise}

Stochastic noise was modeled with \textsc{Discovery}\footnote{\url{https://github.com/nanograv/discovery}} \citep{discovery}, which evaluates the pulsar-timing likelihood with the Gaussian-process coefficients marginalized analytically and samples the remaining hyperparameters with the No-U-Turn sampler (NUTS) as implemented in \textsc{NumPyro} \citep{numpyro}. The linearized timing model was included as a further Gaussian process and marginalized with an effectively unbounded prior; the noise hyperparameters reported below account for their covariance with the timing model.

The white noise of each backend $b$ was defined by rescaling the reported ToA uncertainty $\sigma_i$ as
\begin{equation}
  \sigma_{i}^{2}\;\longrightarrow\;\mathcal{E}_{b}^{2}\,\sigma_{i}^{2}\;+\;\mathcal{Q}_{b}^{2},
  \label{eq:white}
\end{equation}
where $\mathcal{E}_{b}$ (EFAC) rescales the formal uncertainty and $\mathcal{Q}_{b}$ (EQUAD) is added in quadrature, outside EFAC, following the TempoNest convention \citep{Temponest}. Where included, ECORR was implemented as an additional term correlated between the sub-banded ToAs of a single epoch and uncorrelated between epochs. EFAC was fitted for every backend in all models, whereas EQUAD and ECORR were tested but, as set out below, do not form part of the adopted model.

Achromatic red noise (RN) and DM noise were both modeled as a Gaussian process with a Fourier basis of $C$ frequencies $f_{k}=k/T$ ($k=1\ldots C$, with $T$ the data span) and a power-law power spectral density, i.e.
\begin{equation}
  P(f) \;=\; \frac{A^{2}}{12\pi^{2}}\left(\frac{f}{f_{\rm yr}}\right)^{-\gamma}\,\mathrm{yr}^{3}.
  \label{eq:powerlaw}
\end{equation}
where $A$ is the amplitude, $\gamma$ the spectral index, and $f_{\mathrm{yr}}$ a reference frequency. The achromatic basis is frequency-independent, while the DM basis carries the cold-plasma scaling $(\nu_{\rm ref}/\nu)^{2}$ with $\nu_{\rm ref}=1400$\,MHz. The DM amplitude $A_{\rm DM}$ quoted below is therefore the amplitude of the timing delay produced by the DM process at 1400\,MHz. A solar-wind Gaussian process, using the same power-law form on a basis scaled by the dispersive delay of a $1/r^{2}$ solar-wind density model, was also tested.

The adopted noise model was written to a \texttt{TEMPO2} parameter file. The spectral index was carried across unchanged, as \texttt{TNDMGam}, and the number of Fourier components as \texttt{TNDMC}. The DM amplitude required conversion, because $A_{\rm DM}$ is a timing delay referenced to 1400\,MHz whereas the \texttt{TNDMAmp} of the TempoNest convention is expressed in dispersion-measure units and its power spectrum does not carry the factor $1/12\pi^{2}$ of Eq~(\ref{eq:powerlaw}). The two are related by
\begin{equation}
\begin{aligned}
  \mathtt{TNDMAmp}
  &= \log_{10}A_{\rm DM}
     + \log_{10}\!\left(
       \frac{K\,\nu_{\rm ref}^{2}}{\sqrt{12\pi^{2}}}
       \right) \\
  &= \log_{10}A_{\rm DM} + 1.6375 .
\end{aligned}
\label{eq:tndm}
\end{equation}
where $K=2.41\times10^{-4}$\,pc\,cm$^{-3}$\,MHz$^{-2}$\,s$^{-1}$ is the dispersion constant, so that $K\nu_{\rm ref}^{2}=472.36$ and $\sqrt{12\pi^{2}}=10.883$. In our case, for the adopted model this gives $\mathtt{TNDMAmp}=-10.25$.

Both Gaussian processes used $C=88$ components, corresponding to a shortest sampled period of $\sim$110\,d.  A Gaussian process with $C$ components carries $2C$ basis functions, and with $N_{\rm epoch}=176$ distinct observing epochs the data constrain at most of order one basis function per epoch, giving $C\lesssim N_{\rm epoch}/2=88$. Above this limit the process is unconstrained between observing epochs, and because the backends occupy largely disjoint intervals in time it can absorb per-backend white noise. This behaviour was verified by repeating the DM-only fit with $C=20,\,45,\,88,\,160$ and $321$ components. Below $C=88$ the recovered DM amplitude and spectral index vary systematically, the spectral index being biased towards steeper values when the basis cannot reach short enough periods. From $C=88$ upward both are stable to within their uncertainties, while the gain in likelihood per added basis function falls by more than an order of magnitude across the same boundary. Information criteria cannot be used to select $C$, since they count only the two power-law hyperparameters per process and assign no cost to the basis, so that larger $C$ always yields a better score.

All priors are uniform over the ranges given in Tab.~\ref{tab:priors}. Posteriors were inspected for truncation against these bounds; where truncation occurred the bound was widened and the model re-run; the  reported results  are free of prior railing. Each model was sampled with NUTS using four independent chains of 2048 posterior draws, following 1000 warm-up iterations per chain, with a maximum tree depth of 10 and a target acceptance probability of 0.9. This yields 8192 posterior samples per model.

\begin{table}
  \centering
  \caption{Uniform prior ranges used in the noise analysis. $\mathcal{Q}$ and ECORR are given in $\log_{10}$(seconds); the amplitudes $A$ are dimensionless in the convention of Eq~(\ref{eq:powerlaw}).}
  \renewcommand{\arraystretch}{1.3} 
  \label{tab:priors}
  \begin{tabular}{ll}
    \hline
    \hline
    Parameter & Prior range \\
    \hline
    $\mathcal{E}_{b}$ (EFAC)                  & $U(0.1,\,5)$     \\
    $\log_{10}\mathcal{Q}_{b}$ (EQUAD)        & $U(-10,\,-4)$    \\
    $\log_{10}$ ECORR$_{b}$                   & $U(-10,\,-5)$    \\
    $\log_{10}A_{\rm RN}$                     & $U(-20,\,-9)$    \\
    $\gamma_{\rm RN}$                         & $U(0,\,7)$       \\
    $\log_{10}A_{\rm DM}$                     & $U(-20,\,-9)$    \\
    $\gamma_{\rm DM}$                         & $U(0,\,7)$       \\
    \hline
  \end{tabular}
\end{table}

Seventeen noise models were compared, formed from the combinations of per-backend EFAC, EQUAD, ECORR, and the achromatic RN, DM and solar-wind Gaussian processes. Models were ranked by the Bayesian information criterion. Since NUTS does not return the marginal likelihood, the information criteria used here are computed from the largest log-likelihood among the scored posterior draws rather than from an optimised maximum likelihood or a Bayesian evidence, so differences of a few units are indicative rather than decisive.

The preferred model consists of per-backend EFAC together with a DM process, and has 13 free parameters. The DM process is strongly required: replacing it with an achromatic red-noise process degrades the BIC by 51, and adding achromatic red noise to it degrades the BIC by 9. No additional white-noise term is favored and adding a solar-wind process yields no gain in likelihood at all.

The parameters of the adopted noise model are listed in Tab.~\ref{tab:noise} and a corner plot with the posterior distributions is provided in Fig.~\ref{fig:noise_corner} in appendix \ref{appendix:C}. The DM process has $\gamma_{\rm DM}=2.34$ and $\log_{10}A_{\rm DM}=-11.89$, the latter corresponding to a peak-to-peak DM variation of approximately $0.15$\,cm$^{-3}$\,pc over the data span. The DM spectral index is consistent, within its 95 per cent credible interval, with the value $\gamma=8/3$ expected for a Kolmogorov turbulent spectrum.

\begin{table}
  \centering
  \caption{Adopted noise model for PSR~J1807$-$2459B.
  Values are posterior medians with 68 per cent credible intervals.}
  \label{tab:noise}
  \renewcommand{\arraystretch}{1.4} 
  \begin{tabular}{@{}lc@{}}
    \hline
    \hline
    Parameter & Value \\
    \hline
    \vspace{2pt}
    $\mathcal{E}_{\rm BCPM2}$                     & $1.18^{+0.30}_{-0.21}$ \\[2pt]
    $\mathcal{E}_{\rm CO\text{-}GUPPI\_LBAND}$    & $0.98^{+0.05}_{-0.05}$ \\[2pt]
    $\mathcal{E}_{\rm GUPPI\_L}$                  & $1.03^{+0.03}_{-0.02}$ \\[2pt]
    $\mathcal{E}_{\rm GUPPI\_S}$                  & $1.02^{+0.03}_{-0.03}$ \\[2pt]
    $\mathcal{E}_{\rm INCO\text{-}GUPPI\_SBAND}$  & $0.90^{+0.08}_{-0.07}$ \\[2pt]
    $\mathcal{E}_{\rm VEGAS\_L}$                  & $1.37^{+0.05}_{-0.05}$ \\[2pt]
    $\mathcal{E}_{\rm VEGAS\_S}$                  & $1.04^{+0.03}_{-0.03}$ \\[2pt]
    $\mathcal{E}_{\rm mkt\_L}$                    & $1.05^{+0.07}_{-0.06}$ \\[2pt]
    $\mathcal{E}_{\rm mkt\_S}$                    & $1.01^{+0.09}_{-0.08}$ \\[2pt]
    $\mathcal{E}_{\rm pks\_afb}$                  & $1.22^{+0.07}_{-0.06}$ \\[2pt]
    $\mathcal{E}_{\rm pks\_dfb4}$                 & $2.06^{+0.54}_{-0.36}$ \\[2pt]
    $\log_{10} A_{\rm DM}$                        & $-11.89^{+0.07}_{-0.07}$ \\[2pt]
    $\gamma_{\rm DM}$                             & $2.34^{+0.22}_{-0.20}$ \\[3pt]

    \hline
  \end{tabular}
\end{table}

Applying this noise model in \texttt{TEMPO2} and refitting the timing model yields a reduced $\chi^{2}$ of 0.994 and a weighted rms residual of 8.9\,$\mu$s. The post-fit residuals, before and after subtraction of the maximum-likelihood realization of the DM Gaussian process, are shown in Fig.~\ref{fig:residuals}. It is clear that the addition of the noise information strongly improves the quality of the timing solution.

\vspace{0.5cm}

\subsection{Results of the timing analysis}

The results of the \texttt{TEMPO2/Discovery} fit are shown in Tab.\,\ref{tab:preliminary} for both the DDH and the DDGR binary models. In the latter, the PK parameters are computed from the total and companion masses assuming GR. The DDGR solution is also needed to measure $\dot x$, which could not be fitted in the DDH model because of its strong correlation with $\gamma$ (Sec.~\ref{sec:xdot}). The two solutions describe the data equally well, with the same reduced $\chi^2$ and weighted rms residual.

\begin{figure*}[htbp]
    \centering
    
    \begin{minipage}{0.7\textwidth}
        \centering
        \includegraphics[width=\linewidth]{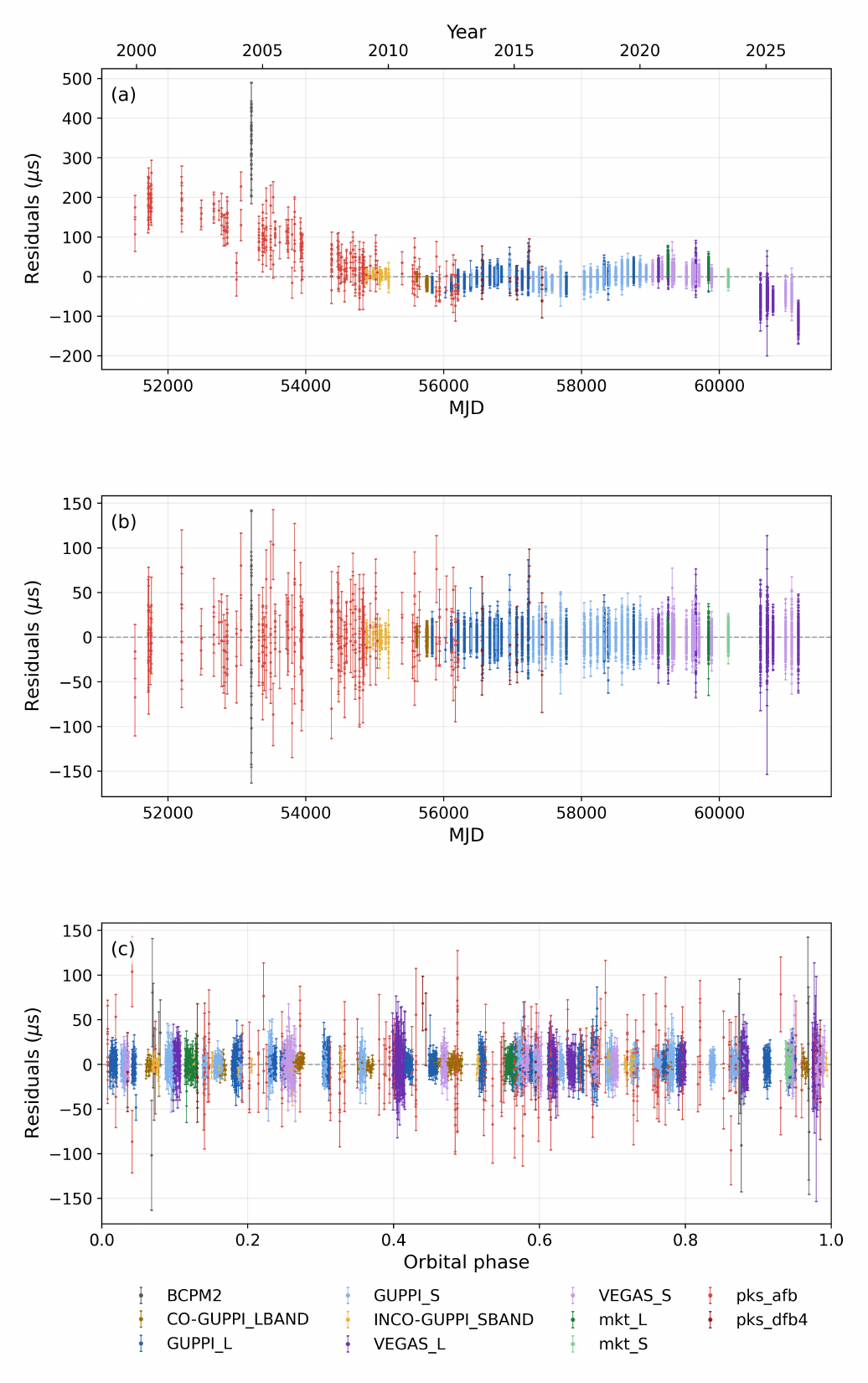}
        \vspace{5mm}
    \end{minipage}
    

    \caption{Post-fit timing residuals of PSR J1807$-$2459B as a function of time and orbital phase using the DDH binary model (see Tab.\,\ref{tab:preliminary}), as returned by \texttt{TEMPO2}. The different receiver and backend combinations mentioned are denoted with different colors. (a) Pre-Noise Model residuals vs. MJD. (b) Noise model subtracted residuals vs MJD. (c) Noise model subtracted residuals vs. orbital phase, the latter computed following the  \texttt{TEMPO2}  \texttt{{binphase}} convention. Notice that this phase includes the contributions from the orbital-period derivative and the fitted periastron advance, and therefore does not correspond directly to the mean anomaly.}
    \label{fig:residuals}
\end{figure*}

The timing residuals are shown in Fig.\,\ref{fig:residuals} as a function of time and orbital phase, before and after subtracting the noise model. After the subtraction, the residuals display no observable trends, which implies that the DDH solution provides, within the measurement uncertainties, an adequate description of the observed ToAs. The comparison with the residuals before the subtraction also shows the importance of modeling the DM variations for this pulsar.

We measured all the PK parameters of the DDH model with high significance: $\dot\omega$, $\gamma$, the two Shapiro-delay parameters $h_3$ and $\varsigma$ and, for the first time, $\dot P_{\rm b}$, which is dominated by the acceleration of the binary in the cluster potential (Sec.~\ref{sec:pbdot}). In particular, the statistical uncertainty of $\dot\omega$ is 7 times smaller than that of the literature value, $0.018319(12)^\circ$\,yr$^{-1}$, currently available on the PSRCAT (v.\,2.9.0) catalogue\footnote{\url{https://www.atnf.csiro.au/research/pulsar/psrcat/}} \citep{PSRCAT_2005}. Note that the uncertainties in Tab.\,\ref{tab:preliminary} are statistical only: the systematic contributions due to the proper motion, which affect $\dot\omega$ and $\gamma$, are discussed in Sec.~\ref{sec:discussion}. We also detect a significant second derivative of the spin frequency, as is common for pulsars in globular clusters \citep{Freire_2017_F23}), while the third derivative is consistent with zero. 

Thanks to our long baseline, we also obtained a good measurement of the total proper motion (PM), $\mu_{\rm tot}=19.1(2.4)$\,mas\,yr$^{-1}$, consistent with that of the cluster, 18.71(6)\,mas\,yr$^{-1}$ \citep{Vasiliev_2019}. Overall, our results are consistent with those of \cite{lynch2012} and improve the precision of several parameters, in particular $\dot\omega$ and $\gamma$. All the PK parameters are discussed in detail in Sec.~\ref{sec:discussion}.

\begin{table*}
\caption[]{Parameters for PSR J1807$-$2459B from the \texttt{TEMPO2/Discovery} analysis, for the DDH and DDGR binary models, and derived quantities. Numbers in parentheses are the nominal 1$\sigma$ statistical uncertainties on the last quoted digits. \textit{(a)} White-noise parameters are reported in Tab.\,\ref{tab:noise}. \textit{(b)} Computed after subtracting the realization of the DM noise obtained in the fit. \textit{(c)} DM variations were modeled using a Taylor expansion about the reference epoch, in addition to the stochastic DM noise; the quadratic component of the DM-noise realization is included in the Taylor coefficients. \textit{(d)} Not fitted in the DDGR model: computed assuming GR from the fitted $M_{\rm tot}$ and $m_{2}$ and the Keplerian parameters; uncertainties propagated from those of $M_{\rm tot}$ and $m_{2}$, assumed uncorrelated. \textit{(e)} For the DDGR solution, $\dot P_{\rm b}$ is fitted as an excess term in addition to the orbital decay due to gravitational-wave emission, $\dot P_{\rm b}^{\rm GR} = -2.79\times10^{-14}$\,s\,s$^{-1}$, computed internally assuming GR from the masses. \textit{(f)} Value derived using the measure of $\dot{\omega}$. \textit{(g)} Distances estimated using the NE2001 and YMW16 Galactic models.}
\label{tab:preliminary}
\centering

\footnotesize{
\renewcommand{\arraystretch}{1.18}
\begin{tabular}{lcc}

\hline
\hline
\rule{0pt}{8pt}Binary model  & DDH & DDGR \\
\hline
\multicolumn{3}{c}{Observation and data reduction parameters} \\
\hline
\rule{0pt}{10pt}Span of timing data (MJD) \dotfill & \multicolumn{2}{c}{51521 -- 61143} \\
Total time span (yr) \dotfill & \multicolumn{2}{c}{26.34} \\
Number of ToAs \dotfill & \multicolumn{2}{c}{3093} \\
Reference epoch (MJD) \dotfill & \multicolumn{2}{c}{56332} \\
Time scale \dotfill & \multicolumn{2}{c}{TDB} \\
Solar system ephemeris \dotfill & \multicolumn{2}{c}{DE440} \\
Weighted rms residual ($\mu$s) \dotfill & 8.92$^{(b)}$ & 8.92$^{(b)}$ \\
Reduced $\chi^2$ \dotfill & 0.994 & 0.994 \\

\hline
\multicolumn{3}{c}{Noise model fit parameters$^{(a)}$} \\
\hline
\rule{0pt}{10pt}DM noise amplitude, $\log_{10}A_{\rm DM}$ \dotfill & \multicolumn{2}{c}{$-11.89$} \\
DM noise spectral index, $\gamma_{\rm DM}$ \dotfill & \multicolumn{2}{c}{$2.34$} \\

\hline
\multicolumn{3}{c}{Astrometric, spin and dispersion parameters} \\
\hline
\rule{0pt}{8pt}Right ascension, $\alpha$ (J2000) \dotfill & 18:07:20.871353(54) & 18:07:20.871353(54) \\
Declination, $\delta$ (J2000) \dotfill & $-$25:00:01.882(22) & $-$25:00:01.883(22) \\
Proper motion in $\alpha$, $\mu_\alpha \cos{\delta}$ (mas yr$^{-1}$) \dotfill & $-$1.904(77) & $-$1.904(76) \\
Proper motion in $\delta$, $\mu_\delta$ (mas yr$^{-1}$) \dotfill & $-$19.0(2.5) & $-$19.1(2.4) \\
Spin frequency, $\nu$ (Hz) \dotfill &  238.881433963346(15) & 238.881433963346(14) \\
Spin frequency first derivative, $\dot \nu$ ($10^{-15}$ Hz s$^{-1}$) \dotfill & $-$4.7002(1)   & $-$4.7002(1) \\
Spin frequency second derivative, $\ddot \nu$ ($10^{-26}$ Hz s$^{-2}$) \dotfill & $-$0.73(8)   & $-$0.72(8) \\
Spin frequency third derivative, $\dddot \nu$ ($10^{-35}$ Hz s$^{-3}$) \dotfill & 0.2(9) & 0.2(9) \\
Dispersion measure, DM (cm$^{-3}$ pc) \dotfill & 137.1418(92)$^{(c)}$ & 137.1418(91)$^{(c)}$ \\
Dispersion measure first derivative, DM1 (cm$^{-3}$ pc yr$^{-1}$) \dotfill & 0.0050(26)$^{(c)}$ & 0.0049(26)$^{(c)}$ \\
Dispersion measure second derivative, DM2 (cm$^{-3}$ pc yr$^{-2}$) \dotfill & 0.00039(41)$^{(c)}$ & 0.00040(41)$^{(c)}$ \\

\hline
\multicolumn{3}{c}{Keplerian orbital parameters} \\
\hline
\rule{0pt}{8pt}Orbital period, $P_{\rm b}$ (days) \dotfill & 9.95666799776(88) & 9.95666799781(87) \\
Projected semi-major axis of the pulsar orbit, $x$ (lt-s) \dotfill & 28.9216(25) & 28.920382(37) \\
Epoch of periastron, $T_0$ (MJD) \dotfill & 54881.34717581(14) & 54881.34717577(12) \\
Orbital eccentricity, $e$ \dotfill & 0.747033167(89) & 0.747033213(30) \\
Longitude of periastron at $T_0$, $\omega$ ($^\circ$) \dotfill & 11.33415(99) & 11.334615(14) \\

\hline
\multicolumn{3}{c}{Post-Keplerian orbital parameters} \\
\hline
\rule{0pt}{8pt}Rate of advance of periastron, $\dot \omega$ ($^\circ$ yr$^{-1}$) \dotfill & 0.0183328(17) & 0.0183336(5)$^{(d)}$ \\
Orbital period derivative, $\dot P_{\rm b}$ ($10^{-11}$ s s$^{-1}$) \dotfill & 1.533(33) & 1.537(32)$^{(e)}$ \\
Rate of change of projected semi-major axis, $\dot x$ ($10^{-15}$ lt-s s$^{-1}$) \dotfill & -- & 2.0(5.1) \\
Orthometric amplitude of the Shapiro delay, $h_3$ ($\mu$s) \dotfill & 4.58(33) & 4.44(18)$^{(d)}$ \\
Orthometric ratio of the Shapiro delay, $\varsigma$ \dotfill & 0.898(20) & 0.907(12)$^{(d)}$ \\
Einstein delay, $\gamma$ (s) \dotfill & 0.0136(17) & 0.014424(25)$^{(d)}$ \\

\hline
\multicolumn{3}{c}{Mass and inclination parameters} \\
\hline
\rule{0pt}{8pt}Mass function, $f$ (M$_\odot$) \dotfill & 0.262011(68) & 0.26197917(99) \\
Total mass, $M_{\rm tot}$ (M$_\odot$) \dotfill & 2.57062(37)$^{(f)}$ & 2.57079(11) \\
Companion mass, $m_{2}$ (M$_\odot$) \dotfill & -- & 1.2065(16) \\
Pulsar mass, $m_{1}$ (M$_\odot$) \dotfill & -- & 1.3643(16)$^{(d)}$ \\
Sine of the orbital inclination, $\sin i$ \dotfill & -- & 0.9952(13)$^{(d)}$ \\

\hline
\multicolumn{3}{c}{Derived parameters} \\
\hline
\rule{0pt}{8pt}Galactic longitude, $l$ (deg) \dotfill & \multicolumn{2}{c}{5.836} \\
Galactic latitude, $b$ (deg) \dotfill & \multicolumn{2}{c}{$-$2.206} \\
Offset from cluster center (core radii) \dotfill  & \multicolumn{2}{c}{4.05} \\ 
Total transverse proper motion, $\mu_{\rm tot}$ (mas yr$^{-1}$) \dotfill & 19.1(2.4) & 19.2(2.4) \\
Position angle of proper motion (J2000), $\Theta_\mu$ ($^\circ$) \dotfill & 185.73(77) & 185.70(76) \\
DM-derived distance (NE2001), $d$ (kpc) \dotfill & \multicolumn{2}{c}{2.94$^{(g)}$} \\
DM-derived distance (YMW16), $d$ (kpc) \dotfill & \multicolumn{2}{c}{3.04$^{(g)}$} \\
Spin period, $P$ (ms) \dotfill &  4.18617714825607(26) & 4.18617714825607(26) \\
Apparent spin period derivative, $\dot P$ ($10^{-20}$ s s$^{-1}$) \dotfill & 8.23671(24)& 8.23671(24) \\
Intrinsic spin period derivative, $\dot P_\mathrm{int}$ ($10^{-20}$ s s$^{-1}$) \dotfill & $0.77(16)$ & $0.76(16)$ \\
Spin-down power, $\dot E$ ($10^{33}$ erg s$^{-1}$) \dotfill & $4.1(9)$ & $4.1(8)$ \\
Surface magnetic field strength, $B_\mathrm{surf}$ ($10^8$ G) \dotfill & $1.81(19)$ & $1.80(19)$ \\
Characteristic age, $\tau_c$ (Gyr) \dotfill & $8.7^{+2.3}_{-1.5}$ & $8.8^{+2.3}_{-1.5}$ \\
\hline


\end{tabular}
}
\end{table*}

\section{Discussion}\label{sec:discussion}


In the previous work on this system, \cite{lynch2012} detected the relativistic advance of periastron, the Shapiro delay and a poorly constrained Einstein delay parameter.
A Shapiro-only analysis (with the DDH model) provided $\sin i\approx0.99715(20)$ and $m_2\approx1.02(17)\,\msun$. 
Using the DDGR timing model, which self-consistently takes into account the Shapiro delay and the tight constraint on the total mass introduced by the measurement of the rate of advance of periastron ($\dot{\omega}$), they derived a companion mass of $m_2=1.2064(20)\,\msun$ and a pulsar mass of $m_1=1.3655(21)\,\msun$. This and the lack of eclipses and orbital variability that would be caused by a main-sequence companion led them to conclude that the companion is either a massive white dwarf or another neutron star. This implies that the components of the PSR~J1807$-$2459B system behave as point masses, making the system well-suited for testing General Relativity. In order to do that, we need to disentangle all of the possible  contributions that can contaminate the  measurement of the post-Keplerian parameters.   

\subsection{Rate of change of the orbital period}
\label{sec:pbdot}
In general, the observed spin-period derivative of a pulsar, $\dot{P}_{\mathrm{obs}}$, differs from its intrinsic value, $\dot{P}_{\mathrm{int}}$, because of several kinematic and gravitational effects along the line of sight, which can be calculated differentiating the expresion for the first-order Doppler shift:
\begin{equation}
\left(\dfrac{\dot{P}}{P}\right)_{\mathrm{obs}} = \left(\dfrac{\dot{P}}{P}\right)_\mathrm{int}  + \frac{a_{\mu}}{c}  + \frac{a_{\mathrm{Gal}}}{c} + \frac{a_{\mathrm{GC}}}{c} ,
\label{pobs}
\end{equation}
where $a_\mu$
represents the so--called Shklovskii contribution due to the proper motion of the system, $a_{\mathrm{Gal}}$ is the relative line-of-sight acceleration between the pulsar and Solar System barycenter due to the Galactic potential, $a_{\mathrm{GC}}$ is the  acceleration of the pulsar along the line of sight due to the cluster gravitational potential (and possibly nearby stars) and $c$ is the speed of light.

We now estimate the acceleration terms by the order they appear.

\subsubsection{The Shklovskii contribution}

The Shklovskii acceleration is given by \citep{Shklovskii_1970}:
\begin{equation}
a_\mu = 0.7283\times 10^{-12}\left( \dfrac{d}{\rm kpc} \right) \left( \dfrac{\mu_{\rm tot}}{\rm mas\; yr^{-1}} \right)^2    
\end{equation}
where we use $d=2580$ pc and $\mu_{\rm tot}=18.7$ mas yr$^{-1}$, i.e. the most recent values quoted in literature for the cluster \citep{Vasiliev_2019}, which is consistent with our timing measurement. We therefore estimate an acceleration of $a_\mu = 6.57 \times 10^{-10}$ m s$^{-2}$ and  a contribution to  $\dot{P}_b$ of $1.89 \times 10^{-12}$ s s$^{-1}$. This value would have been  $2.19 \times 10^{-12}$ s s$^{-1}$ if we had assumed a distance of 3 kpc as reported in \cite{lynch2012}.

\subsubsection{The Galactic acceleration}

We estimated the Milky Way acceleration imparted to the system
by applying Eq.\,(16) in \cite{lazaridis}, also using for the vertical component of the Galactic acceleration $F_z$ the analytic formula provided by \citet[their Eq.\,(14)]{widrow_Gaia}. The adopted values for the Solar motion in the Galaxy are $\Theta_\odot =240.5\pm4.1$\,km\,s$^{-1}$ and $R_\odot = 8.275\pm0.034$\,kpc \citep{gravita_2021}. The resulting acceleration is $a_{\rm Gal}\simeq 1.0 \times 10^{-10} \mathrm{m\,s^{-2}}$, where we used the galactic coordinates $l=5.84$° and $b=-2.2$°. This implies a contribution to $\dot{P}_b$ of  $2.8 \times 10^{-13}$ s s$^{-1}$.  Notice that for this system the vertical contribution is almost negligible, given the small galactic latitude. If we employ the simpler model provided in  \cite{prager_2017}, we obtain a slightly different value, namely $a_{\rm Gal}\simeq 1.1 \times 10^{-10}\, \rm m / s^2$
and hence a contribution to $\dot{P}_b$ of about $3.3 \times 10^{-13}$\, \rm s\, s$^{-1}$.

\subsubsection{The globular cluster acceleration}

From the observed orbital period derivative, we can subtract the terms associated with the first two contributions to derive $a_{\rm GC, psr} = (4.60 \pm 0.12) \times 10^{-9} $ m s$^{-2}$. 
In dense globular clusters such as NGC 6544, $a_{\mathrm{GC}}$ is generally, and by far, the dominant part. This means that pulsars in the far side of the cluster, which have negative line-of-sight components of their accelerations, can have a negative $\dot{P}_\mathrm{obs}$, as in the case of PSR~J1807$-$2459A, which has $\dot P_{\mathrm{obs}}=-4.3352(26)\times10^{-21}$ s s$^{-1}$ \citep{lynch2012}.

We can estimate the maximum contribution with the approximate formula \citep{Phinney1993,lynch2012} 
\begin{equation}
   \dfrac{a_{\rm GC,max}}{c} \approx \dfrac{3 \sigma^2_v}{2c \sqrt{r_c^2+r_{\perp}^2}}
\end{equation}
where $\sigma_v$ is the central velocity dispersion, $r_c$ is the core radius and $r_{\perp}$ is the projected distance of the pulsar from the center of the cluster. For NGC\;6544, we have $\sigma_v=5.89$ km/s, $r_c=0.05$'  $\simeq 0.037$ pc and $r_{\perp}=4.05\; r_c \simeq 0.15$ pc, which gives $a_\mathrm{GC,max} \approx 1.08 \times 10^{-8}$ m s$^{-2}$. This is larger than $a_{\rm GC,psr}$, i.e., the latter is within the expected range of cluster accelerations.

A more precise estimate for the cluster acceleration requires a model for the cluster mass. Indeed, the line-of-sight component of the acceleration imparted by the cluster can be expressed as
\begin{equation}
  a_{\rm GC}=-GM(r_{\rm psr})
  \frac{r_\parallel}{(r_{\perp}^2+r_\parallel^{2})^{3/2}},
\label{eq:accel}
\end{equation}
where $G$ is Newton's gravitational constant, $M(r_{\rm psr})$ is the
cluster mass enclosed in a radius equal to $r_{\rm psr}$ (i.e. the distance from the cluster center), and $r_{\perp}$ and $r_\parallel$ are the pulsar coordinates in the cluster frame, perpendicular and parallel to the line of sight, respectively. Even assuming $|r_\parallel| \approx |r_\perp|$ (which returns $r_{\rm psr}\approx 0.215$ pc), the equation above is not informative without an accurate estimate for the mass profile $M(r)$. For NGC\;6544, which is classified as a core-collapsed globular cluster \footnote{The cluster NGC\;6544 is highly concentrated and dynamically evolved, exhibiting properties consistent with a core-collapsed or post-collapse state. For such systems, the central surface brightness profile often exhibits a distinct cusp, with no evidence for a flat core, as demonstrated observationally in systems such as M15
\citep{sosins_1997} and discussed in the context of
globular cluster evolution \citep{Baumgardt_2018,Vasiliev_2019}.},
adopting a specific analytical mass profile $M(r)$ (e.g. a standard King model \citep{King_1962}) may introduce significant systematic uncertainties in the estimate of the enclosed mass at the pulsar location. However, it is possible to derive the mass profile from N-body fitting model for NGC\;6544 \citep{Baumgardt_2023}. 
Using best fitting data for the N-body model, the cumulative mass profile of NGC\;6544 is described by the following numerical fit 
\begin{equation}
M(r) \simeq 10^{\rm P [\log_{10} (r/{\rm pc}) ]} \, \msun,
\end{equation}
where
\begin{equation}
P(y)=0.043\,y^4-0.040\,y^3-0.407\,y^2+0.936\,y+4.337
\end{equation}
valid in the range of radial distance covered by data, namely  $0.1 \, \rm pc \le r \le 30 \, \rm pc$. Using this profile, we computed the expected line-of-sight acceleration $a_{\rm GC}(u)$ as a function of the unknown geometrical parameter $u \equiv r_\parallel/r_\perp$. 

\begin{figure}[htbp]
    \centering
    \includegraphics[width=0.999\columnwidth]{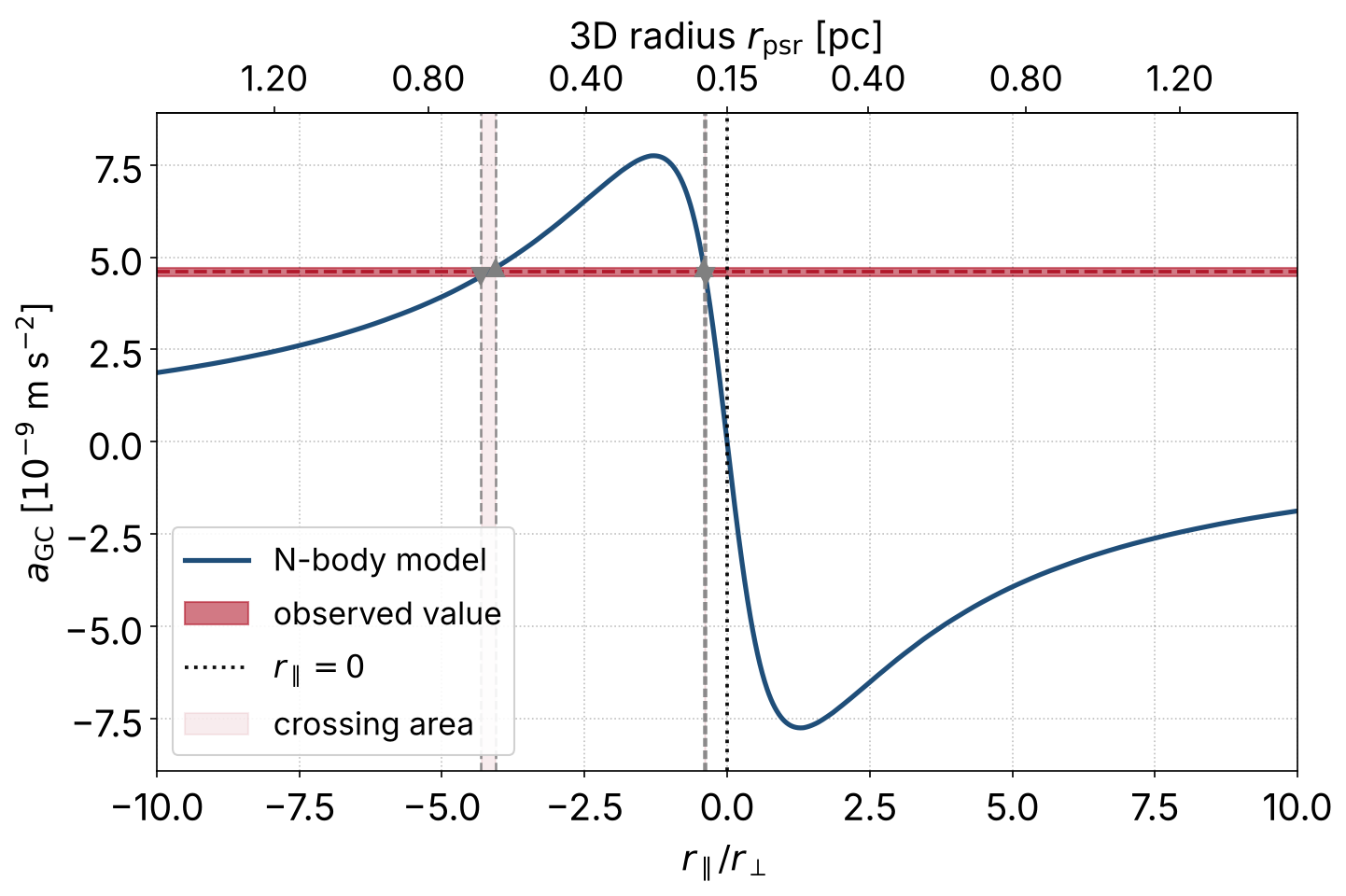} 
    \caption{Predicted line-of-sight gravitational acceleration $a_{\rm GC}$ as a
function of $r_\parallel/r_\perp$,
computed from the cumulative mass profile $M(r)$ of the best-fitting
N-body model of NGC 6544. The horizontal shaded region represents the observed
acceleration interval $a_{\rm GC}=(4.60\pm0.12)\times10^{-9}$\,m\,s$^{-2}$,
while the vertical dashed lines mark the intersections between the model
prediction and the observed $1\sigma$ band. The upper horizontal axis shows
the corresponding three-dimensional distance of the pulsar from the
cluster center.}
    \label{fig:accel}
\end{figure}

The predicted $a_{\rm GC}$ for the line of sight of PSR~J1807$-$2459B ($r_\perp = 4.05 \, r_c$) is shown in Fig.\;\ref{fig:accel}. It reaches a maximum value of $\sim 7.7\times10^{-9}\,\mathrm{m\,s^{-2}}$ at $u\simeq-1.29$, again exceeding $a_{\rm GC, psr}$. As a consequence, the observed $1\sigma$ interval of $a_{\rm GC, psr}$ intersects the model prediction in two disconnected regions of parameter space. The first allowed solution corresponds to relatively small line-of-sight displacements, $u\in[-0.40,\,-0.38]$, implying $r_\parallel\in[-0.061,\,-0.057]\,\mathrm{pc}$. The second allowed solution corresponds instead to larger displacements, namely $u\in[-4.31,\,-4.05]$, corresponding to $r_\parallel\in[-0.65,\,-0.61]\,\mathrm{pc}$.

These results therefore place strong constraints on the allowed three-dimensional
location of the pulsar within the cluster, exactly $r_{\rm psr} \simeq 0.160\text{--}0.162~\mathrm{pc}$ or $r_{\rm psr} \simeq 0.63\text{--}0.66~\mathrm{pc}$. Notice that $r_{\rm psr} < r_h$ for both solutions. The positive line-of-sight acceleration places PSR J1807$-$2459B on the near side of the cluster; within the adopted N-body mass model, it lies either very close to the plane of the sky through the cluster center at $r_{\rm psr}\simeq0.16$ pc or $\sim$0.6 pc in front of it.  
A complementary, model-independent constraint on the enclosed mass is presented in Appendix~\ref{appendix:A}.


\subsubsection{Intrinsic spin-down of the pulsar}

The intrinsic spin period derivative $\dot P_{\rm int}$ can be derived without computing the acceleration terms: since these terms are the same in Eq.~(\ref{pobs}) and in its analogue for $P_{\rm b}$, subtracting the two relations and assuming $\dot P_{\rm b,int}=\dot P_{\rm b}^{\rm (GR)}$ gives
\begin{equation}
\dot P_{\rm int}=\dot P_{\rm obs}-P\,
\frac{\dot P_{\rm b,obs}-\dot P_{\rm b}^{\rm (GR)}}{P_{\rm b}}.
\label{eq:pdotint}
\end{equation}
The advantage of this expression is that it is independent on the distance, the proper motion, the Galactic
potential or the position of the pulsar in the cluster. The uncertainty
is dominated by that of $\dot P_{\rm b}$. Using the values in Tab.~\ref{tab:preliminary}, we obtain
$\dot P_{\rm int}=(7.7\pm1.6)\times10^{-21}$ s\,s$^{-1}$, and hence $\tau_c \simeq 8.7$ Gyr (below the age of the cluster) and $B_{\rm surf} \simeq 1.8 \times 10^{8}$ G. Therefore, once the kinematic contribution measured through $\dot P_{\rm b}$ is removed, PSR J1807$-$2459B turns out to be a typical, fully recycled millisecond pulsar.

\subsection{Rate of change of the length of the projected semi-major axis and Einstein delay}
\label{sec:xdot}

In the DDH solution we did not fit for $\dot x$. For this system the Einstein delay can be separated from the Rømer delay only through the slow rotation of the orbit caused by the periastron advance ($\Delta\omega\simeq0.48^\circ$ over our 26.3-yr baseline), and the resulting secular signature is indistinguishable from that of a change in $x$  \citep{Ridolfi_2019_correlation}. The small value of $\omega$ makes this worse. Fitting $\gamma$ and $\dot x$ simultaneously therefore yields an almost singular solution, and in the DDH fit $\dot x$ was held fixed at zero. In the DDGR model, where $\gamma$ is not a free parameter, $\dot x$ can be measured: we obtain $\dot x=(2.0\pm5.1)\times10^{-15}$ lt-s s$^{-1}$.

A $\dot{x}$ could arise due to a number of physical and geometric contributions, which can be decomposed as \citep{book_pulsar_2004}:
\begin{equation}
\dot{x}_{\mathrm{obs}} = \dot{x}_\mu + \dot{x}_{\rm \dot{D}} + \dot{x}_{\rm GR} + \dot{x}_{\dot{\rm M}} + \dot{x}_{\rm 3^{rd}} +  \dot{x}_{\dot{\epsilon}_{\rm A}} + \dot{x}^{\rm P}_{\rm SO} + \dot{x}^{\rm C}_{\rm SO}.
\end{equation}
We now describe each term and estimate its magnitude. 

\subsubsection{Evaluation of contributions to $\dot{x}$}

The first term, $\dot{x}_\mu$ is caused by the proper motion of the system. It is given by \citep{kopeikin_1996}
\begin{equation}
\label{eqn:xdot_pm}
\dot{x}_\mu \simeq  x \mu_{\rm tot} \cot{(i)} \sin(\Theta_{\mu} - \Omega) 
\end{equation}
where $\Theta_{\mu}$ is the position angle of the proper motion vector,  $\Omega$ the ascending node angle, and $i$ the orbital inclination. Here, $\mu_{\rm tot}$ is in units of rad s$^{-1}$. Using $\mu_{\rm tot}=18.7$ mas yr$^{-1}$ (Sec.~\ref{sec:pbdot}) and $\sin i=0.9952(13)$ from the DDGR solution, we obtain $|\dot x_\mu|\le x\mu_{\rm tot}|\cot i|\simeq8.2\times10^{-15}$ lt-s s$^{-1}$. The DDGR value of $\dot x$ is consistent with this bound, but its uncertainty is comparable to the whole allowed range [$\sin(\Theta_\mu-\Omega)=0.24\pm0.62$], so it does not constrain $\Omega$.

The second term, $\dot{x}_{\rm \dot{D}}$, is due to the changing radial Doppler shift, $\cal{A}_{\rm l}$, which includes the pulsar's acceleration in the gravitational field of the cluster. The third term, $ \dot{x}_{\rm GR}$, is caused by the shrinkage of the orbit caused by gravitational wave emission. The fourth term, $\dot{x}_{\dot{\rm M}}$, is caused by possible mass loss in the system. These three effects contribute to both $\dot{P}_{b}$ and $\dot{x}$. Since we have a measurement of $\dot{P}_{b}$, the contributions from these effects to both the measurements can be related, to within the same order of magnitude, as (see e.g. \citealt{Damour1992_tests_of_gravity}) 
\begin{equation}
    \label{eqn:xdot_pbdot}
    \left(\frac{\dot{x}}{x}\right) \sim  \left(\frac{\dot{P}_{\rm b}}{P_{\rm b}}\right).
\end{equation}
From this, we obtain a total corresponding contribution from these terms more than one order of magnitude lower than the proper motion contribution, namely $\simeq 5 \times 10^{-16}$ lt-s s$^{-1}$. Notice that the only GR contribution is $\dot{x}_{\rm GR}=-6.3 \times 10^{-19}$ lt-s s$^{-1}$, which is orders of magnitude smaller than the proper motion term. 

The fifth term, $\dot{x}_{\rm 3^{rd}}$ is caused by the presence of a hypothetical third body in the system. Even if we cannot exclude the presence of a third-body, this term is usually negligible compared to others. Moreover,  the presence of a third body would induce significant drifts in the timing of the pulsar that usually require the addition of several higher-order spin derivatives and we find no evidence for spin derivatives higher than $\dddot{\nu}$ with the latter being consistent with zero and not crucial for our final solution.
The magnitudes of $\dot{\nu}$ and $\ddot{\nu}$ are normal for pulsars in GCs and can be attributed to the pulsar's motion through the cluster's gravitational potential.

The $\dot{x}_{\dot{\epsilon}_{\rm A}}$ term arises from the secular change of the aberration parameter $\epsilon_A$ caused by geodetic precession of the pulsar spin. From Eqs.~(2.6), (3.20), (3.24), (3.34) and (3.35) of \cite{Damour1992_tests_of_gravity}, its magnitude is bounded by $|\dot{x}_{\dot\epsilon_A}| \le x\,(P/P_b)\,\Omega_1^{\rm spin}/(\sin i\sqrt{1-e^2}\,\sin^2\lambda) \simeq 6\times10^{-19}/\sin^2\lambda$ lt-s s$^{-1}$, where $\Omega_1^{\rm spin} \simeq 5.1\times10^{-3}$ deg yr$^{-1}$ and $\lambda$ is the angle between the spin axis and the line of sight. The term vanishes if the spin is aligned with the orbital angular momentum. Since PSR J1807-2459B is likely an exchange product, alignment cannot be assumed. The term is largest for a nearly perpendicular spin, but it remains below $\sim10^{-17}$ lt-s s$^{-1}$ unless the spin axis lies within a few degrees of the line of sight.  Therefore it is safely negligible.

The remaining two terms, $\dot{x}^{\rm P}_{\rm SO}$ and $ \dot{x}^{\rm C}_{\rm SO}$,  result from the spin-orbit coupling from the fast spin of the pulsar (P) and the companion (C).  These contributions vanish for a spin aligned with the orbital angular momentum and are maximal for a spin along the line of nodes. Since PSR J1807-2459B is likely an exchange product where any spin-orbit configuration is possible, we take the latter case to derive an upper limit. By using Eqs. (3.7), (3.20) and (3.27) in \cite{Damour1992_tests_of_gravity} and assuming for the pulsar $I_{\rm 1} = (1$–$1.5)\times10^{45}$ g cm$^{2}$ we obtain $|\dot x^{\rm P}_{\rm SO}| \le (6$–$8)\times10^{-16}$ lt-s s$^{-1}$, about an order of magnitude below both our measurement uncertainty and $|\dot x_\mu|_{\max}$.  For a neutron-star companion, $|\dot x^{\rm C}_{\rm SO}| \lesssim 10^{-15}$ lt-s s$^{-1}$ even if it were itself recycled. For a massive white-dwarf companion, the contribution $|\dot x^{\rm C}_{\rm SO}|$ would exceed our uncertainty only for spin periods $\lesssim 30$ s, which is unlikely for a white dwarf acquired in an exchange encounter.

In conclusion, all contributions except for the one due to proper motion are negligible, and thus $\dot{x}_{\rm obs} \approx \dot{x}_\mu$.

\subsubsection{Einstein delay}

\cite{lynch2012} measured $\gamma=26(14)$\,ms, consistent with GR but significant only at the $1.9\sigma$ level, and anticipated that a more precise measurement would provide a second test of GR. Our value ($\gamma = 13.6 \pm 1.7 \, \rm ms$) is consistent with theirs ($0.9\sigma$) and 8 times more precise: the Einstein delay is now detected at $8\sigma$ level.

However, our measurement of $\gamma$ cannot be interpreted as in compact relativistic binaries, where it arises almost entirely from the variation of the second-order Doppler shift and the gravitational redshift. 
For wide binaries like PSR J1807$-$2459B there is a correlation between $\dot{x}$ and $\gamma$ which can be quantitatively investigated by computing the effect of $ \gamma$ on $\dot{x}$ through  \citep{Ridolfi_2019_correlation}
\begin{equation}\label{eq:xdot_gamma}
\dot{x}_{\gamma} = -\frac{\gamma \,\dot{\omega}}{\sqrt{1-e^2}} \sin\omega  
\end{equation}
which evaluates to $-4 \times 10^{-14}$ lt-s s$^{-1}$ in our case. This term should be added to other contributions if one adopts a timing model that does not account for the Einstein delay, as discussed in \cite{Guo_2021A&}, which is not the case of the DDH model.  In any case, this quantity needs to be compared with the proper motion contribution. Since $|\dot x_\mu|_{\max}\simeq0.2\,|\dot x_\gamma|$, the kinematic contribution is not negligible: any value of $\dot x$ within its allowed range is absorbed by the DDH fit as a change of $\gamma$, and, as a result, the value of $\gamma$ in Tab.\,\ref{tab:preliminary} needs to be corrected by a systematic error.  Repeating the DDH fit with $\dot x$ fixed at $\pm|\dot x_\mu|_{\max}$ gives $\gamma=$ 16.34  and 10.92 ms respectively (with unchanged $\chi^2$), as expected from Eq.~(\ref{eq:xdot_gamma}). We therefore adopt half of this difference as a systematic uncertainty, $\gamma=13.6\pm1.7\,({\rm stat})\pm 2.7\,({\rm sys})$ ms, i.e. $\gamma=13.6\pm3.2$ ms when combined in quadrature. This means that our actual measurement is only 4.3-$\sigma$ significant.

\subsection{Rate of advance of periastron}\label{omdot_section}

The orbital eccentricity of the system and the long timing baseline allow a highly significant measurement of the rate of advance of periastron, despite the wide orbit. Our DDH solution gives $\dot{\omega} =0.0183328(17)$$^\circ$ yr$^{-1}$. In this model $\dot{\omega}$ is strongly correlated with $\gamma$ ($\rho\simeq0.95$).   

\subsubsection{Proper motion and other contributions}

The observed $\dot{\omega}$ is expected to be almost entirely due to the relativistic periastron precession. In order to gauge the reliability and meaning of the measurement of $\dot{\omega}$, we first consider the possibility of additional non-relativistic effects. Given the high proper motion of PSR~J1807$-$2459B, the most important of these additional contributions is given by \citep{kopeikin_1996}
\begin{equation}\label{pm_omdot_equation}
	\dot{\omega}_\mu = \frac{\mu_{\rm tot}}{\sin{i}} \cos(\Theta_\mu - \Omega),
\end{equation}
Here, $\mu_{\rm tot}$ is in units of deg yr$^{-1}$. Assuming an optimal alignment ($\cos(\Theta_\mu - \Omega)=1$) and $\sin{i}=0.9952$ (from the DDGR solution), it contributes at the order of $\dot{\omega}_\mu \simeq 5.2 \times 10^{-6}$ deg yr$^{-1}$, which is about three times larger than our statistical uncertainty and therefore cannot be neglected. As a result, we treat this contamination as an additional source of systematic uncertainty. Notice that this contribution was not considered in \cite{lynch2012}.

In addition, since $\dot x$ is held fixed at zero in our DDH solution and $\dot\omega$ is strongly correlated with $\gamma$, the unknown kinematic $\dot x_\mu$ (Sec.~\ref{sec:xdot}) also biases the fitted $\dot\omega$, by $\delta\dot\omega\approx-(\dot x_\mu/x)\cot\omega=-\mu_{\rm tot}\cot i\cot\omega\,\sin(\Theta_\mu-\Omega)$. Our refits with $\dot x=\pm8.2\times10^{-15}$ lt-s s$^{-1}$ give $\delta\dot\omega=+2.57\times10^{-6}$ and $-2.49\times10^{-6}$ deg yr$^{-1}$, in agreement with this expression. This term is negligible  when $\omega$ is close to $90^\circ$ because it scales as $\cos i\cot\omega$ relative to Eq.~(\ref{pm_omdot_equation}); however, here $\omega\simeq11^\circ$ gives $\cos i\cot\omega\simeq0.5$. 

Since the two proper-motion terms depend on $\Omega$ as a cosine and a sine, respectively, their \textit{maximum} combined effect is their sum in quadrature; adding also the statistical uncertainty, we obtain $\dot\omega=0.0183328\pm0.0000017\,({\rm stat})\pm0.0000058\,({\rm sys})$ deg yr$^{-1}$, i.e. $\dot\omega=0.0183328(60)$ deg yr$^{-1}$ when the two are combined in quadrature.

A third body in the system can  give an additional contribution to the observed periastron advance, but, as discussed in Sec.~\ref{sec:xdot}, we have no evidence for a third body, so we do not consider it.  Other contributions to $\dot\omega$, namely higher-order relativistic terms, spin--orbit coupling and a tidal or rotationally induced quadrupole of the
companion (see also \citealt{lynch2012}), are negligible compared with our total uncertainty, unless the companion is a white dwarf rotating with a period of about a minute or less, which is unlikely for a companion acquired in an exchange encounter. 

\subsubsection{Relativistic periastron advance}

In GR, the leading order term of the relativistic periastron advance depends on only one unknown quantity, the total mass of the system $M_\mathrm{tot}$:
\begin{equation}\label{eq:mtot}
	M_\mathrm{tot} = \frac{1}{T_\odot} \left[\frac{\dot{\omega}}{3}(1-e^2)\right]^{3/2} \left(\frac{P_\mathrm{b}}{2\pi}\right)^{5/2},
\end{equation}
where $T_\odot\equiv {\cal GM}_\odot^\mathrm{N}/c^3= 4.92549094764126\, \rm \mu s$ is an exact quantity that follows from the exact definitions of the speed of light $c$ and the solar mass parameter ${\cal GM}_\odot^\mathrm{N}$ \citep{Psra_2016A}. From it, we derive a total mass  of  $M_{\rm tot}=2.57062 \pm 0.00037\, ({\rm stat}) \pm 0.0012\, ({\rm sys})\, \msun$, i.e. $M_{\rm tot} = 2.5706 \pm 0.0013 \, M_{\odot}$. Applying Eq.\,(\ref{eq:mtot}) to the DD solution of \cite{lynch2012} yields $M_{\rm tot}=2.5677\pm0.0025~\, \msun$; therefore, the two measurements are statistically consistent. In both cases, the uncertainty in the total mass is almost entirely driven by the uncertainty in $\dot{\omega}$.

Overall, our statistical uncertainty on $\dot\omega$ is a factor of 7 smaller than that of \cite{lynch2012} (DD solution); including the proper-motion systematics, which were neglected there, the overall improvement is a factor of 2, and so is the improvement in $M_{\rm tot}$ ($0.0013$ vs. $0.0025\,{\rm M_\odot}$).

\subsection{Constraints on masses and orbital geometry} 
\label{sec:masses}

\begin{figure*}[t]
\centering
 \includegraphics[width=\hsize]{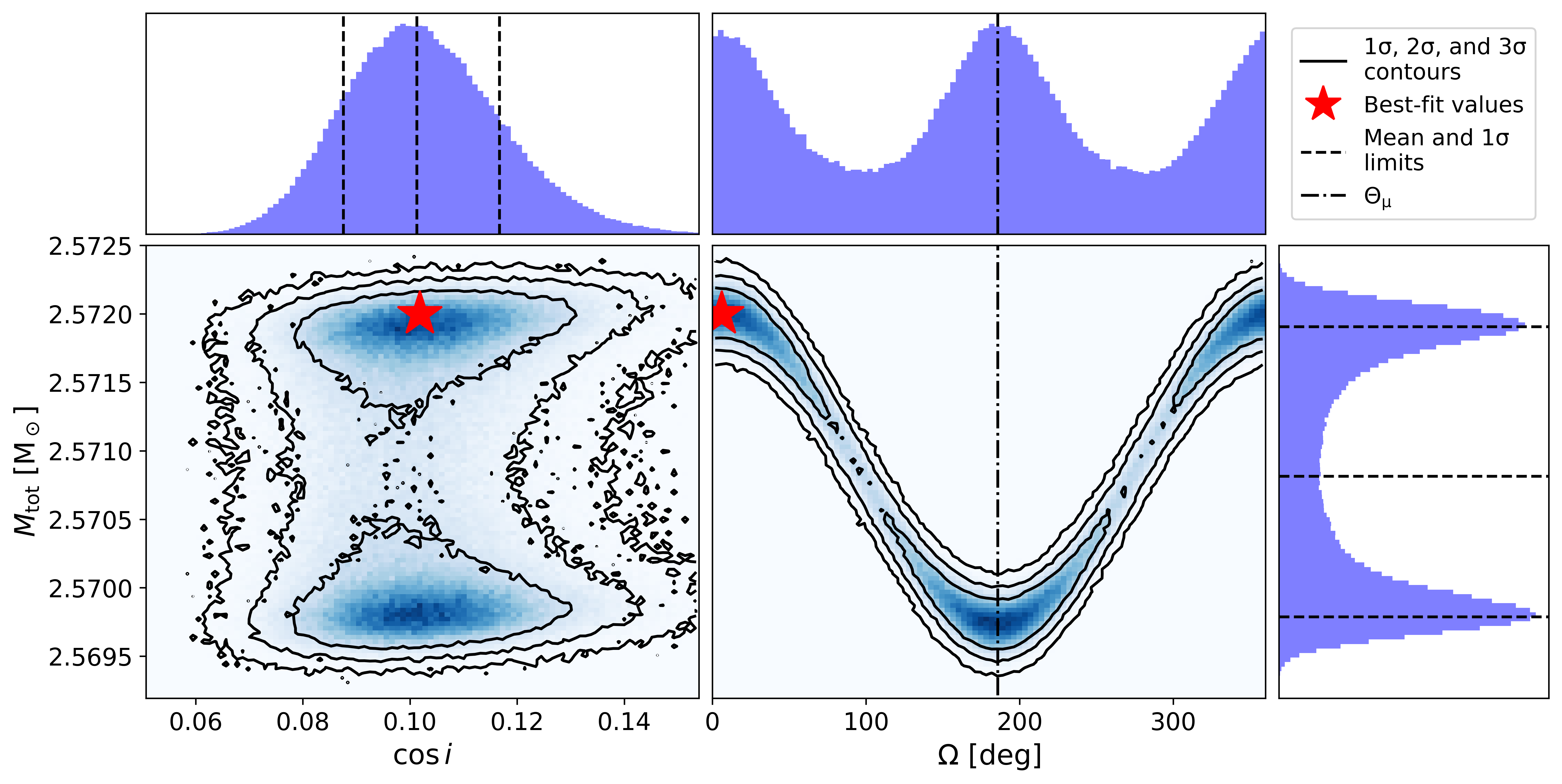}
 \includegraphics[width=\hsize]{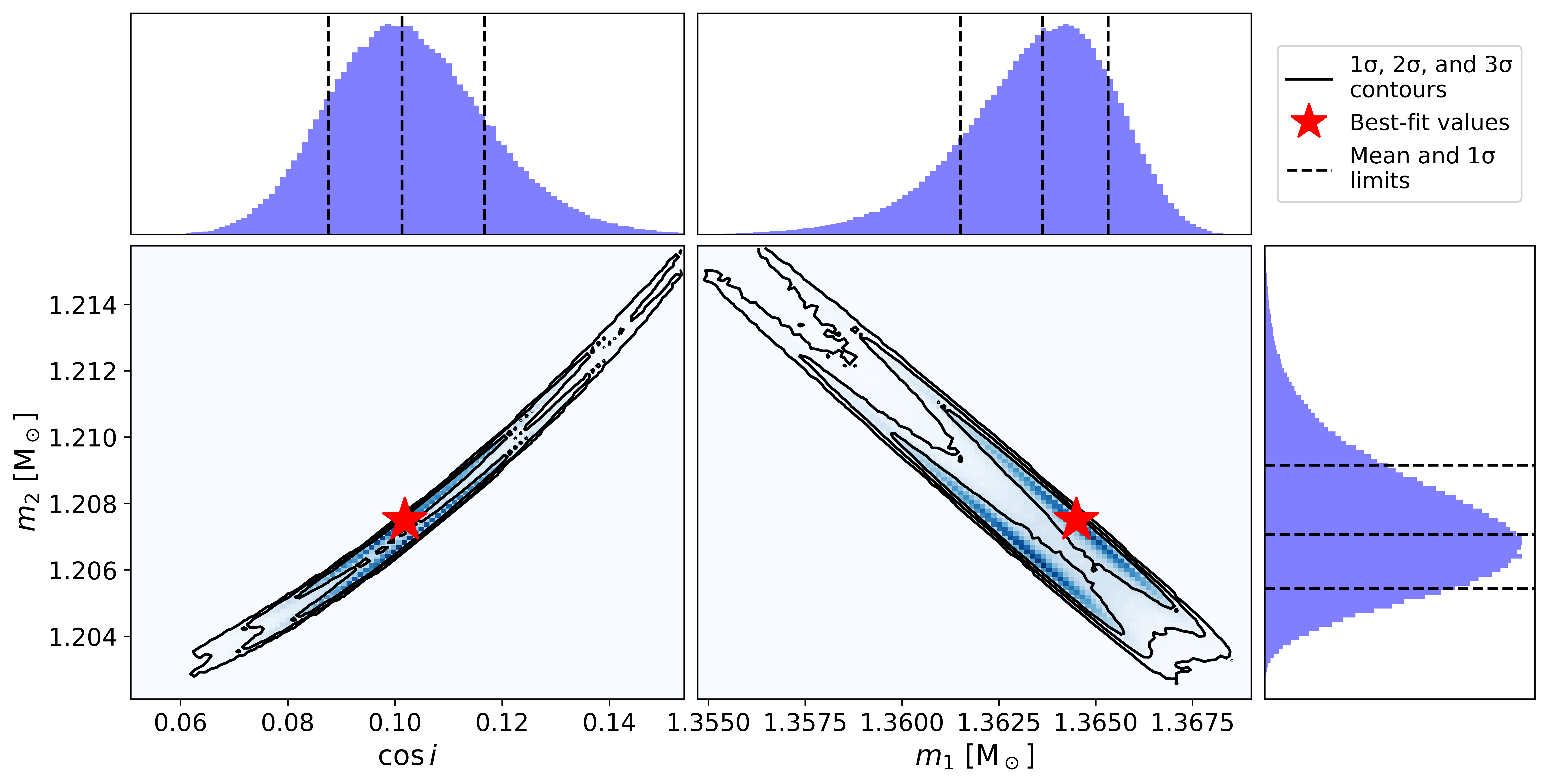}
 \caption{Density diagrams derived from 2\,174\,400 \texttt{emcee} samples. Top: constraints on $M_{\rm tot}$, $|\!\cos{i}|$, and $\Omega$. Bottom: the same for $m_1$, $m_2$, and $|\cos{i}|$, where $m_1$ and $m_2$ are derived using the mass function. Samples from the $\cos{i}>0$ and $\cos{i}<0$ runs have been added together with their corresponding weights. The best-fit point corresponds to the highest sampled likelihood.}
 \label{mass-diagrams}
\end{figure*}

\begin{figure*}[t]
\centering
 \includegraphics[width=\textwidth]{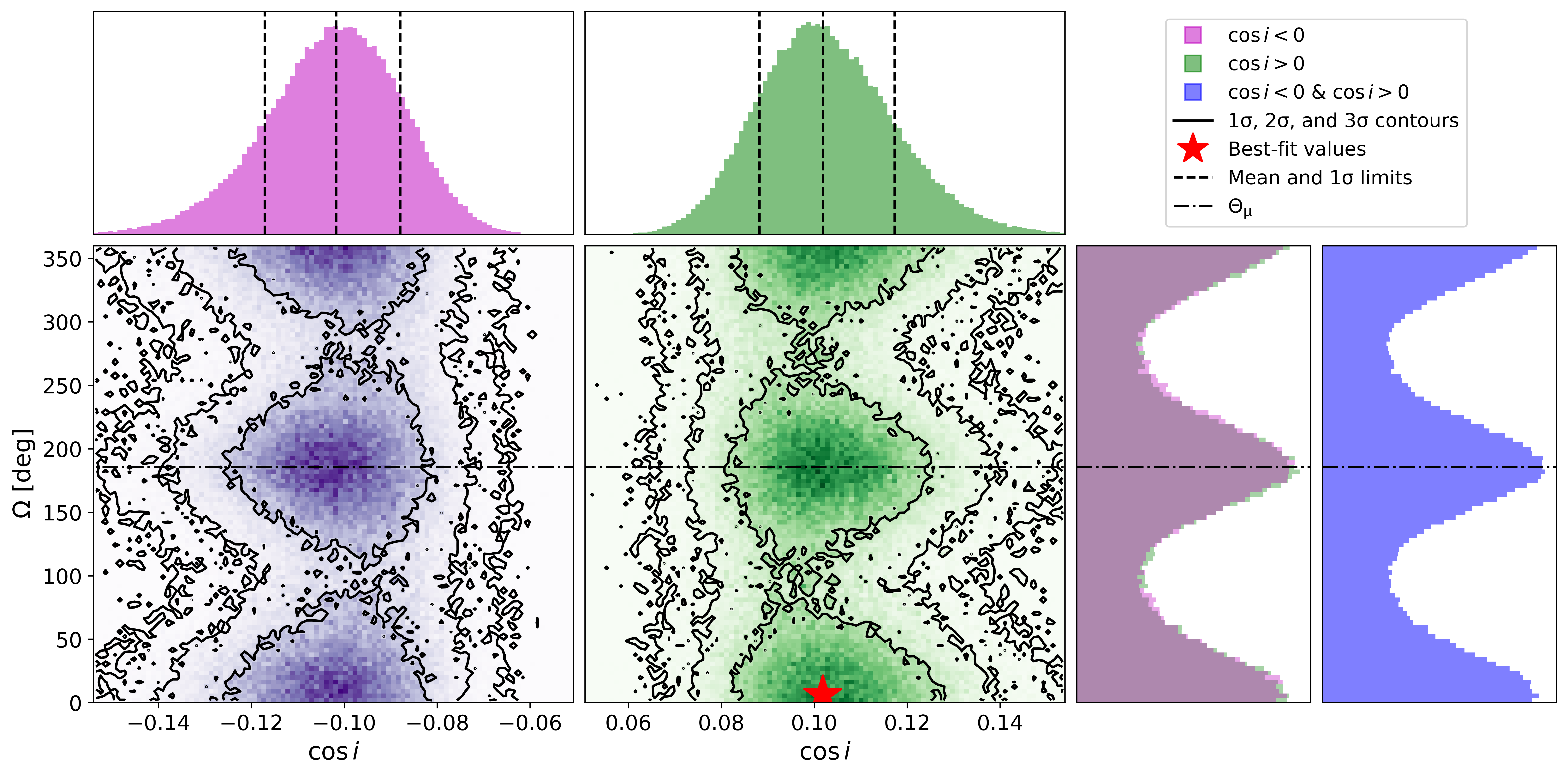}
 \caption{Density diagrams on the $\cos{i}$--$\Omega$ space, resolving the $\cos{i}>0$ and $\cos{i}<0$ \texttt{emcee} runs with 1\,087\,200 samples on each side. On the right-most plot, the corresponding weights have been applied to the samples before addition. The best-fit point corresponds to the highest sampled likelihood.}
 \label{geometry-diagram}
\end{figure*}


We investigated the correlation of $M_{\rm tot}$ with $\Omega$ via Eq.~(\ref{pm_omdot_equation}), the probability distributions of $m_1$, $m_2$, and $\cos{i}$, and the probability distribution of $\Omega$. As is common in timing experiments \citep[e.g.][]{splaver2002probing,pillay2026constraints},
we implemented a Bayesian exploration of the allowed masses and orbital configurations. In our case, we explored the $M_{\rm tot}$--$\cos{i}$--$\Omega$ parameter space, computing the values of several PK parameters as a function of the values from their relevant GR and PM contributions. This includes the rate of periastron advance,
\begin{equation}
	\dot{\omega} = \dot{\omega}_{\rm GR} + \dot{\omega}_\mu,
\end{equation}
the change in the projected semi-major axis,
\begin{equation}
	\dot{x} = \dot{x}_\mu + \dot{x}_{\rm \dot{D}},
\end{equation}
and $\gamma$, $h_3$ and $\varsigma$, which contain only GR contributions.

As described in Sec.~\ref{omdot_section}, the $\dot{\omega}_\mu$ contribution is potentially significant compared to the $\dot{\omega}$ measurement uncertainty, making the inclusion of this effect paramount to understand the true distribution of possible $M_{\rm tot}$ values. This is especially relevant for adequate mass constraints, as DDGR can underestimate the uncertainty of $M_{\rm tot}$ due to not accounting for the $\dot{\omega}_\mu$ contribution. For $\dot{x}$, the $\dot{x}_\mu$ contribution is expected to be dominant and comparable to the measurement uncertainty (Sec.~\ref{sec:xdot}). The $\dot{x}_{\rm \dot{D}}$ contribution is estimated via
\begin{equation}
    \left(\frac{\dot{x}_{\rm \dot{D}}}{x}\right) = \left(\frac{\dot{P}_{\rm b, \dot{D}}}{P_{\rm b}}\right),
\end{equation}                         
where $\dot{P}_{\rm b, \dot{D}}=\dot{P}_{\rm b, obs}-\dot{P}_{\rm b}^{\rm (GR)}$ is the same quantity entering Eq.\,(\ref{eq:pdotint}) and  $\dot{P}_{\rm b, obs}$ is the observed $\dot{P}_{\rm b}$ in the DDH fit. During the Bayesian fit, the resulting $\dot{x}_{\rm \dot{D}}=5.147\times10^{-16}$\,lt-s\,s$^{-1}$ contribution was kept as constant. We ignored the aberration and SO contributions to $\dot{x}$ due to their low significance and the unknown SO orbit configuration, which does not enable an accurate prediction. Finally, $\dot{P}_{\rm b}$ is dominated by the Doppler factor contribution, which does not depend on $M_{\rm tot}$, $\cos{i}$, nor $\Omega$, so we left it as a free parameter.

The Bayesian exploration of the $M_{\rm tot}$--$\cos{i}$--$\Omega$ space was implemented by an \texttt{emcee} \citep{emcee} wrap around \texttt{TEMPO2}. The prior was uniform, and for each $\{M_{\rm tot},\cos{i},\Omega\}$ sample, we predicted and fixed the mentioned PK parameters (except for $\dot{P}_{\rm b}$) in the DDH parameter file. The position and PM parameters were also fixed to their best-fit values to ensure consistency with the proper motion contributions, but all remaining spin, DM, Keplerian parameters, as well as $\dot{P}_{\rm b}$, were left as free. After each fit, we evaluated the $\chi^2$ and the log-likelihood was computed as
\begin{equation}
    \log\!\mathcal{L}(M_{\rm tot},\cos{i},\Omega)=-\frac{1}{2}\chi^2.
\end{equation}

Since the $\dot{x}_\mu$ contribution breaks the sign symmetry in the $\cos{i}$ axis, this implementation also has the potential to result in different $\Omega$ distributions at $\cos{i}<0$ and $\cos{i}>0$, hence enabling multiple valid $\cos{i}-\Omega$ configurations. Therefore, we set two independent runs: one restricted at $\cos{i}<0$, and another restricted at $\cos{i}>0$. Both runs had the same amount of \texttt{emcee} walkers and steps, and the samples were added together for the final analysis. However, to ensure a proper weighting of each half, we computed the accumulated probability on each side via
\begin{equation}
    \mathcal{P}=\sum_{\rm bins}\langle\mathcal{L}(M_{\rm tot},\cos{i},\Omega)\rangle,
\end{equation}
which adds the average $\langle\mathcal{L}\rangle$ value within uniformly spaced bins in the $M_{\rm tot}$--$\cos{i}$--$\Omega$ space (computed from the average $\mathcal{L}$ of the samples within each bin) which are symmetric in the $\cos i < 0$ and $\cos i > 0$ sides. Then, the weights of each run are computed as
\begin{equation}
    \mathcal{W}_{+}=\frac{\mathcal{P}_{+}}{\mathcal{P}_{+}+\mathcal{P}_{-}},\quad\mathcal{W}_{-}=\frac{\mathcal{P}_{-}}{\mathcal{P}_{+}+\mathcal{P}_{-}}.
\end{equation}
These weights were applied to the samples of each run before adding them to infer parameter distributions.

After discarding burn-in, the runs included a total of 2\,174\,400 samples. The resulting weights from the $\cos{i}>0$ and $\cos{i}<0$ runs were $\mathcal{W}_{+}=0.4998$ and $\mathcal{W}_{-}=0.5002$, which yields a Bayes factor of $\mathcal{B}=\mathcal{W}_{+}/\mathcal{W}_{-}=0.9991$, indicative of equal weight between the two $\cos{i}$ sign configurations.

Figure~\ref{mass-diagrams} shows the constraints on $M_{\rm tot}$, $|\!\cos{i}|$ and $\Omega$, as well as the derived parameters $m_1$ and $m_2$ (from the mass function), after the addition of the $\cos{i}>0$ and $\cos{i}<0$ samples with the appropriate weighting. The constrained masses and $1\sigma$ uncertainties are $m_1=1.3636^{+0.0017}_{-0.0021}$\,M$_\odot$, $m_2=1.2071^{+0.0021}_{-0.0017}$\,M$_\odot$, which are consistent with \cite{lynch2012} and show very similar uncertainties. However, unlike the previous analysis, this work accounts for the proper-motion contribution to $\dot\omega$, making our results more realistic. For the inclination angle, we obtained $|\!\cos{i}|=0.102^{+0.015}_{-0.014}$ or $|i|=84.1^{+0.8}_{-0.9}$\,deg. The total system mass is constrained to $M_{\rm tot}=2.5708^{+0.0011}_{-0.0010}$\,M$_\odot$. A full Bayesian analysis of all the timing parameters, in which GR is assumed through a prior on $x$, yields consistent masses (Appendix~\ref{appendix:B}).

The Bayesian exploration also shows the importance of accounting for the proper motion contribution to $\dot{\omega}$. Our derived $M_{\rm tot}$ uncertainty is 10 times higher than that of the bare DDGR, and consistent with the constraint derived in Sec.~\ref{omdot_section}. The exploration of the parameter space also shows a bimodal distribution for $M_{\rm tot}$. This manifests as two parallel, high-probability sequences in the $m_2$--$m_1$ diagram, and as two distinct high-probability regions in the $M_{\rm tot}$--$\cos{i}$ and $M_{\rm tot}$--$\Omega$ diagrams. The latter shows a clear sinusoidal correlation between $M_{\rm tot}$ and $\Omega$ due to the $\dot{\omega}_\mu$ contribution. The probability density within the sinusoid is modulated by the $\dot{x}_\mu$ contribution, resulting in the two distinct regions. The bimodal $M_{\rm tot}$ distribution has two distinct peaks at $M_{\rm tot}=2.5698$\,M$_\odot$ and $M_{\rm tot}=2.5719$\,M$_\odot$.

\begin{table}[t]
\caption[]{\label{likelihood_peaks_table} Parameter values at the likelihood peaks in Figs.~\ref{mass-diagrams} and \ref{geometry-diagram}.}
\centering
\begin{tabular}{ccccc}
\hline
\hline\\[-1.9ex]
 $M_{\rm tot}$ & $\cos{i}$ & $\Omega$ & $\dot{\omega}_\mu$ & $\dot{x}_\mu$ \\
 (M$_\odot$) &  & (deg) & (deg\,yr$^{-1}$) & (lt-s\,s$^{-1}$) \\
\hline\\[-1.9ex]
2.5697 &  0.1017 & 185.69   & $5.3\times10^{-6}$ & $6.1\times10^{-18}$ \\
2.5697 & $-0.1024$ & 185.69 & $5.3\times10^{-6}$ & $-6.1\times10^{-18}$ \\
2.5720 & $-0.1017$ & 5.01   & $-5.3\times10^{-6}$ & $1.1\times10^{-16}$ \\
2.5720 &  0.1019 & 6.11  & $-5.3\times10^{-6}$ & $5.8\times10^{-17}$ \\
\hline
\hline
\end{tabular}
\end{table}

The bimodal distribution of $M_{\rm tot}$ is a direct consequence of which regions are preferred in the $\cos{i}$--$\Omega$ space. 
Figure~\ref{geometry-diagram} shows the four preferred regions in the $\Omega$--$\cos{i}$ diagram, determined by the $\dot{x}_\mu$ contribution. While $\Omega$ is not constrained to high significance due to the large widths of the probability distributions, its two preferred values are $\Omega\approx5$\,deg and $\Omega\approx186$\,deg, separated by 181\,deg. This corresponds to the value of $\Omega$ being parallel or antiparallel to the angle of the proper motion, $\Theta_\mathrm{\mu}=185.7$\,deg (Tab.~\ref{tab:preliminary}). 
Tab.~\ref{likelihood_peaks_table} lists the parameter values, as well as the  derived proper motion contributions, at the likelihood peaks. In line with $\Omega$ and $\Theta_{\rm \mu}$ being parallel or antiparallel, the magnitude of $\dot{\omega}_\mu$ is essentially equal to the maximum amplitude predicted by Eq.~(\ref{pm_omdot_equation}) in all four peaks, while the magnitude of $\dot{x}_\mu$ is below two orders of magnitude of the maximum amplitude predicted by Eq.~(\ref{eqn:xdot_pm}).

\subsection{Tests of General Relativity}
\label{sec:GR}

Our DDH solution provides four PK parameters that can be used to test GR: $\dot\omega$, $\gamma$, $h_3$ and $\varsigma$ (Tab.~\ref{tab:preliminary}). The observed $\dot P_{\rm b}$ cannot be used, since it is dominated by the acceleration of the system in the cluster potential (Sec.~\ref{sec:pbdot}). In GR, each PK parameter is a known function of the Keplerian parameters and of the two component masses: two PK parameters therefore determine the masses, and each additional one provides a test of the theory \citep{Damour1992_tests_of_gravity}. Following \cite{Meng_2025}, we derive the masses from the two PK parameters that constrain them best and express each test as the ratio between the measured value of a third PK parameter and its GR prediction; since GR predicts a ratio of unity, the uncertainty of the ratio gives the precision of the test. For PSR~J1807$-$2459B the two parameters used for the masses are $\dot\omega$ and $\varsigma$: unlike in \cite{Meng_2025}, $\gamma$ cannot be used for this purpose, because its total uncertainty ($\simeq23$ per cent; Sec.~\ref{sec:xdot}) would poorly constrain $m_{2}$. 

Assuming that the measured $\dot\omega$ is entirely relativistic, Eq.~(\ref{eq:mtot}) gives $M_{\rm tot}=2.5706\pm0.0013\,{\rm M_\odot}$, where the uncertainty includes the proper-motion systematics discussed in Sec.~\ref{omdot_section}; higher-order (2PN) contributions to $\dot\omega$ are of relative order $3\times10^{-6}$ and therefore negligible. The orthometric ratio depends only on the orbital inclination, $\varsigma=\sin i/(1+|\cos i|)$ \citep{DDH_model_2010}, and can be inverted as
\begin{equation}
\sin i=\frac{2\varsigma}{1+\varsigma^2}.
\label{eq:sini_stig}
\end{equation}
From $\varsigma=0.898(20)$ we obtain $\sin i=0.9942(24)$, i.e. $i=83.9(1.3)^\circ$. The mass function $f$ then gives the companion mass, $m_{2}=(f M_{\rm tot}^2)^{1/3}/\sin i=1.2077(29)\,{\rm M_\odot}$, and the pulsar mass, $m_{1}=M_{\rm tot}-m_{2}=1.3629(30)\,{\rm M_\odot}$. These uncertainties are dominated by that of $\varsigma$ (the uncertainties of $M_{\rm tot}$ and $f$ contribute only $4\times10^{-4}$ and $1\times10^{-4}\,{\rm M_\odot}$ to that of $m_{2}$), and the values are consistent with those of the DDGR solution  in Tab.~\ref{tab:preliminary} and of the Bayesian analysis in the previous section. \\

As a first test of GR, we exploit our measure of $h_3$. In GR, the orthometric amplitude is $h_3=T_\odot\, m_{2}\,\varsigma^3$ \citep{DDH_model_2010}. Inserting the companion mass derived above, the predicted value depends only on $M_{\rm tot}$, $f$ and $\varsigma$:
\begin{equation}
h_{3,\rm GR}=\frac{T_\odot}{2}\,(fM_{\rm tot}^2)^{1/3}\,\varsigma^2\,(1+\varsigma^2)=4.31\pm0.28\,\mu{\rm s},
\label{eq:h3gr}
\end{equation}
where the uncertainty comes from $\varsigma$. Because $h_{3,\rm GR}$ depends strongly on $\varsigma$, the uncertainties of $h_3$ and $\varsigma$ contribute comparably (7.2 and 6.4 per cent, respectively) to that of the ratio $\mathcal{R}\equiv h_3/h_{3,\rm GR}$, and the correlation between the two measurements (correlation coefficient $\rho=-0.29$ in our fit) must be taken into account. Propagating the uncertainties to first order and using the measured $h_3=4.58(33)\,\mu$s, we obtain

\begin{equation}
\frac{h_3}{h_{3,\rm GR}}=1.06\pm0.12,
\label{eq:h3test}
\end{equation}
consistent with GR within $0.5\sigma$. Equivalently, the companion mass implied by the Shapiro delay alone, $m_{2}=h_3/(T_\odot\varsigma^3)=1.28(14)\,{\rm M_\odot}$, agrees with the value derived above. This is the test performed by \cite{lynch2012}, who obtained $m_{2}=1.02(17)\,{\rm M_\odot}$ from the Shapiro delay alone. \\

As a second test of GR, we exploit our measurement of $\gamma$. With the same masses, GR predicts
\begin{equation}
\gamma_{\rm GR}=e\left(\frac{P_{\rm b}}{2\pi}\right)^{1/3}T_\odot^{2/3}\,\frac{m_{2}\,(M_{\rm tot}+m_{2})}{M_{\rm tot}^{4/3}}=14.44\pm0.05\,{\rm ms}.
\label{eq:gammagr}
\end{equation}
Our DDH measurement is $\gamma=13.6\pm1.7\,({\rm stat})\pm2.7\,({\rm sys})$\,ms, where the second term accounts for the unknown kinematic $\dot x_\mu$ (Sec.~\ref{sec:xdot}). Hence
\begin{equation}
\frac{\gamma}{\gamma_{\rm GR}}=0.94\pm0.12\,({\rm stat})\pm0.19\,({\rm sys})=0.94\pm0.22,
\label{eq:gammatest}
\end{equation}
where in the last step the two uncertainties have been added in quadrature. Here the uncertainty of $\gamma_{\rm GR}$  is negligible compared with that of $\gamma$; for this reason, unlike for $h_3$, the correlations of $\gamma$ with $\varsigma$ and $\dot\omega$  do not affect the result. Our result agrees with GR within $0.3\sigma$. 

\begin{figure}[htbp]
    \centering
    \includegraphics[width=0.999\columnwidth]{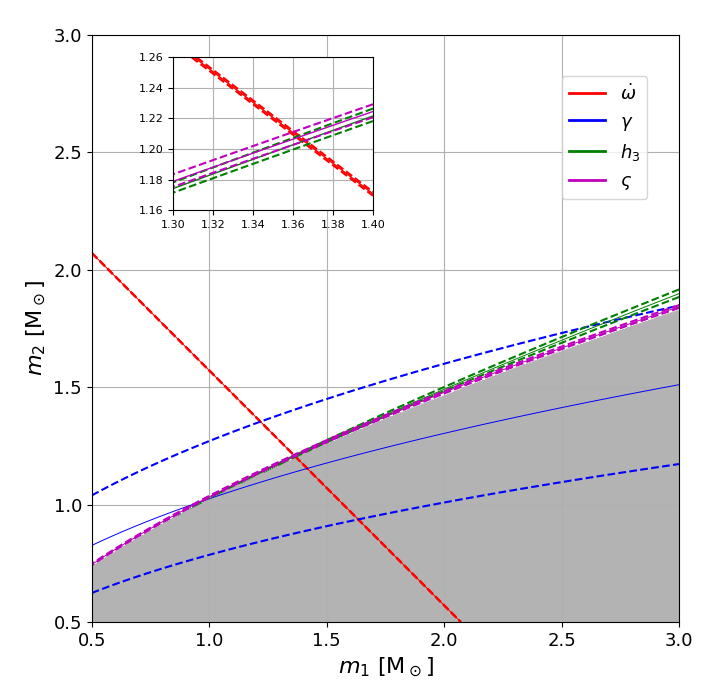} 
    \caption{Mass--mass diagram for PSR~J1807$-$2459B, based on the PK parameters $\dot\omega$, $\gamma$, $h_3$ and $\varsigma$ of our DDH solution (Tab.~\ref{tab:preliminary}), interpreted in GR. For each parameter, the solid line shows the masses for which the GR prediction equals the measured value, and the dashed lines the $1\sigma$ (68.3 per cent) interval; for $\dot\omega$ and $\gamma$ the uncertainties include the proper-motion systematics discussed in Secs.~\ref{sec:xdot} and~\ref{omdot_section}. For $\dot\omega$, the three lines are indistinguishable at this scale. The gray region is excluded by the mass function, since it would require $\sin i>1$. The $h_3$ and $\varsigma$ bands are almost parallel to each other and to the edge of this region: on their own, they mainly constrain the orbital inclination, and the masses are fixed by their intersection with the $\dot\omega$ line. The inset shows this intersection, which lies entirely within the $1\sigma$ band of $\gamma$ (hence not visible in the inset). All four PK parameters are therefore consistent with a common pair of masses, $m_{\rm 1}\simeq1.363\,{\rm M_\odot}$ for the pulsar and $m_{\rm 2}\simeq1.208\,{\rm M_\odot}$ for the companion (Sec.~\ref{sec:GR}). $\dot P_{\rm b}$ is not shown, since it is dominated by the cluster  potential.}
    \label{fig:mass-mass-diagram}
\end{figure}

This 22 per cent test is limited by the systematic term: additional timing will reduce the statistical uncertainty, but the systematic contribution is set by the geometry of the system and will not decrease unless $\Omega$, and hence $\dot x_\mu$, is determined independently. Both tests are summarized in the mass--mass diagram of Fig.~\ref{fig:mass-mass-diagram}, where the regions allowed by $\dot\omega$, $\gamma$, $h_3$ and $\varsigma$, all taken from the DDH solution, overlap in a common region. We notice that these tests probe the quasi-stationary regime of gravity; no test of gravitational-wave damping is possible for this system, since the GR contribution to $\dot P_{\rm b}$ ($-2.8\times10^{-14}$\,s\,s$^{-1}$) is an order of magnitude smaller than the uncertainty of the measured $\dot P_{\rm b}$ and is completely masked by the cluster acceleration.

\section{Summary and conclusions}\label{sec:summary}
\label{sec:end}

We have presented a new timing study of PSR~J1807$-$2459B, a 4.19-ms pulsar in a 9.96-d, highly eccentric ($e\simeq0.75$) orbit around a massive companion, located in the core-collapsed globular cluster NGC\;6544. Our analysis combines more than 26 years of observations (1999--2026) taken with the Parkes, Green Bank and MeerKAT radio telescopes, and includes a model of the dispersion-measure variations, which is essential for this pulsar. The resulting timing solution (Tab.~\ref{tab:preliminary}) improves on that of \cite{lynch2012} and leads to the following main results.

We have measured for the first time the apparent variation of the orbital period, $\dot P_{\rm b}=1.533(33)\times10^{-11}$\,s\,s$^{-1}$. This value is about 600 times larger than, and opposite in sign to, the orbital decay expected from gravitational-wave emission: it is dominated by the acceleration of the binary in the gravitational field of the cluster, $a_{\rm GC,psr}=(4.60\pm0.12)\times10^{-9}$\,m\,s$^{-2}$. This has two consequences. First, the same kinematic contribution also affects the observed spin-down of the pulsar; subtracting it directly we obtain a surface magnetic field of $\simeq1.8\times10^8$\,G and a characteristic age of $\simeq8.7$\,Gyr (below the age of the cluster), so that PSR~J1807$-$2459B turns out to be a typical, fully recycled millisecond pulsar. Second, combined with an N-body mass model of NGC\;6544, the measured acceleration places the pulsar in front of the cluster center, at a 3D distance of either $\simeq0.16$ or $\simeq0.65$\,pc from it.

The rate of advance of periastron is measured with a statistical uncertainty 7 times smaller than before. However, the proper motion of the system adds a contribution that depends on the unknown orientation of the orbit and is about three times larger than the statistical uncertainty. Taking it into account, we obtain a total mass $M_{\rm tot}=2.5706\pm0.0013\,{\rm M_\odot}$, twice as precise as the value implied by the previous measurement of $\dot\omega$. Assuming GR, a Bayesian exploration of the masses and of the orbital orientation gives very precise estimates for a pulsar in a globular cluster, namely a pulsar mass $m_1=1.3636^{+0.0017}_{-0.0021}\,{\rm M_\odot}$, a companion mass $m_2=1.2071^{+0.0021}_{-0.0017}\,{\rm M_\odot}$ and an orbital inclination of $\simeq84^\circ$, consistent with previous studies. 

An optical counterpart is not necessarily expected if the companion is an
old white dwarf. In the standard single-star evolutionary channel, the
progenitor of a $1.2\,M_\odot$ white dwarf would have evolved early, giving
the remnant a cooling age close to the 11--13 Gyr age of the cluster, and
cooling models for such massive white dwarfs \citep[see, e.g.,][]{Camisassa_2019_WD}
predict a very faint object: for the adopted distance and reddening
($(m-M)_0\simeq12$, $E(B-V)\simeq0.8$; \citealt{Cohen_2014}), $V\gtrsim30$ is
plausible, although its precise value depends on the core and atmospheric
composition of the white dwarf and on differential reddening. A recent
merger, on the other hand, could produce a much younger and brighter
remnant. The pulsar lies in a crowded region, $0.2'$ from the cluster
center, and no optical counterpart was found in archival HST/WFC3 images
down to ${\rm F390W}\simeq27.4$ and ${\rm F606W}\simeq26.1$ \citep{Marano_2019_Tesi}, which, according to that analysis, implies a cooling age $\gtrsim0.6$ Gyr for a $1.2\,M_\odot$ white dwarf. Thus, the non-detection is consistent with either an old, cool white dwarf or a neutron star, and only
disfavors a recently formed white dwarf.

We have also measured the Einstein delay, $\gamma=13.6\pm1.7\,({\rm stat})\pm2.7\,({\rm sys})$\,ms, 4.4 times more precisely than \cite{lynch2012}. Because of the wide orbit and of its orientation, $\gamma$ is strongly correlated with a possible change of the projected semi-major axis caused by the proper motion, which we have included as a systematic uncertainty. Together with the Shapiro delay, which we also measure, this allows two independent tests of GR, both passed: $h_3/h_{3,\rm GR}=1.06\pm0.12$ for the Shapiro delay and $\gamma/\gamma_{\rm GR}=0.94\pm0.22$ for the Einstein delay. Continued timing is expected to improve the Shapiro-delay test, whereas the Einstein-delay test is limited by the systematic term, which will not decrease unless the orientation of the orbit is determined independently. Finally, a test of gravitational-wave damping is not possible for this system, since $\dot P_{\rm b}$ is dominated by the cluster acceleration.

PSR~J1807$-$2459B thus shows that a binary pulsar in a dense globular cluster can probe at the same time the gravitational field of the cluster and the predictions of GR, provided that the kinematic effects of its environment and of its proper motion are carefully taken into account. Therefore, this system remains a premier laboratory for the study of fundamental physics and the dynamical evolution of compact binaries in the most crowded regions of our Galaxy.

\section*{Acknowledgements}
Am.C. acknowledges financial support under the National Recovery and Resilience Plan (PNRR), Mission 4, Component 2, Investment 1.1, Call for tender No. 104 published on 2.2.2022 by the Italian Ministry of University and Research (MUR), funded by the European Union - NextGenerationEU - Project Title GUVIRP-Gravity tests in the UltraViolet and InfraRed with Pulsar timing -- CUP C53D23000910006 - Grant Assignment Decree No. 962 adopted on 30/06/2023 by MUR. PF, FK and KG acknowledge ongoing support from the Max-Planck society. FK and KG acknowledge support from the International Max Planck Research School (IMPRS) for Astronomy and Astrophysics at the Universities of Bonn and Cologne. FA acknowledges that part of the research activities described in this paper were carried out with the contribution of the NextGenerationEU funds within PNRR, Mission 4 – Education and Research, Component 2 – From Research to Business (M4C2), Investment Line 3.1 – Strengthening and creation of Research Infrastructures, Project IR0000034 – ‘STILES -Strengthening the Italian Leadership in ELT and SKA’. Part of this work has been funded using resources from the INAF Large Grant 2022 “GCjewels” (P.I. Andrea Possenti) approved with the Presidential Decree 30/2022. JD and DP acknowledge financial support from the INAF Large Grant 2023 "Gravitational wave detection using Pulsar Timing Arrays" (P.I Delphine Perrodin). The National Radio Astronomy Observatory is a facility of the National Science Foundation operated under cooperative agreement by Associated Universities, Inc. The Green Bank Observatory is a facility of the National Science Foundation operated under cooperative agreement by Associated Universities, Inc. SMR is a CIFAR Fellow and is supported by the NSF Physics Frontiers Center award 2020265,  NSF award 2607948, and NSF AAG award 2510064. Pulsar research at UBC is supported by an NSERC Discovery Grant and by the Canadian Institute for Advanced Research. Am.C. is sincerely grateful to Holger Baumgardt for providing the cumulative mass profile data for NGC\;6544.

\bibliographystyle{aa}
\bibliography{references}

\appendix

\section{A}
\label{appendix:A}
For completeness, we considered a purely geometrical, model-independent constraint based only on the measured line-of-sight acceleration and the projected pulsar offset from the cluster center. Apart from spherical symmetry, this construction does not assume any specific mass profile for NGC\;6544 and therefore provides a conservative lower bound on the enclosed mass, together with the associated $M$--$u$ degeneracy. 

\begin{figure}[htbp]
    \centering
    \includegraphics[width=0.999\columnwidth]{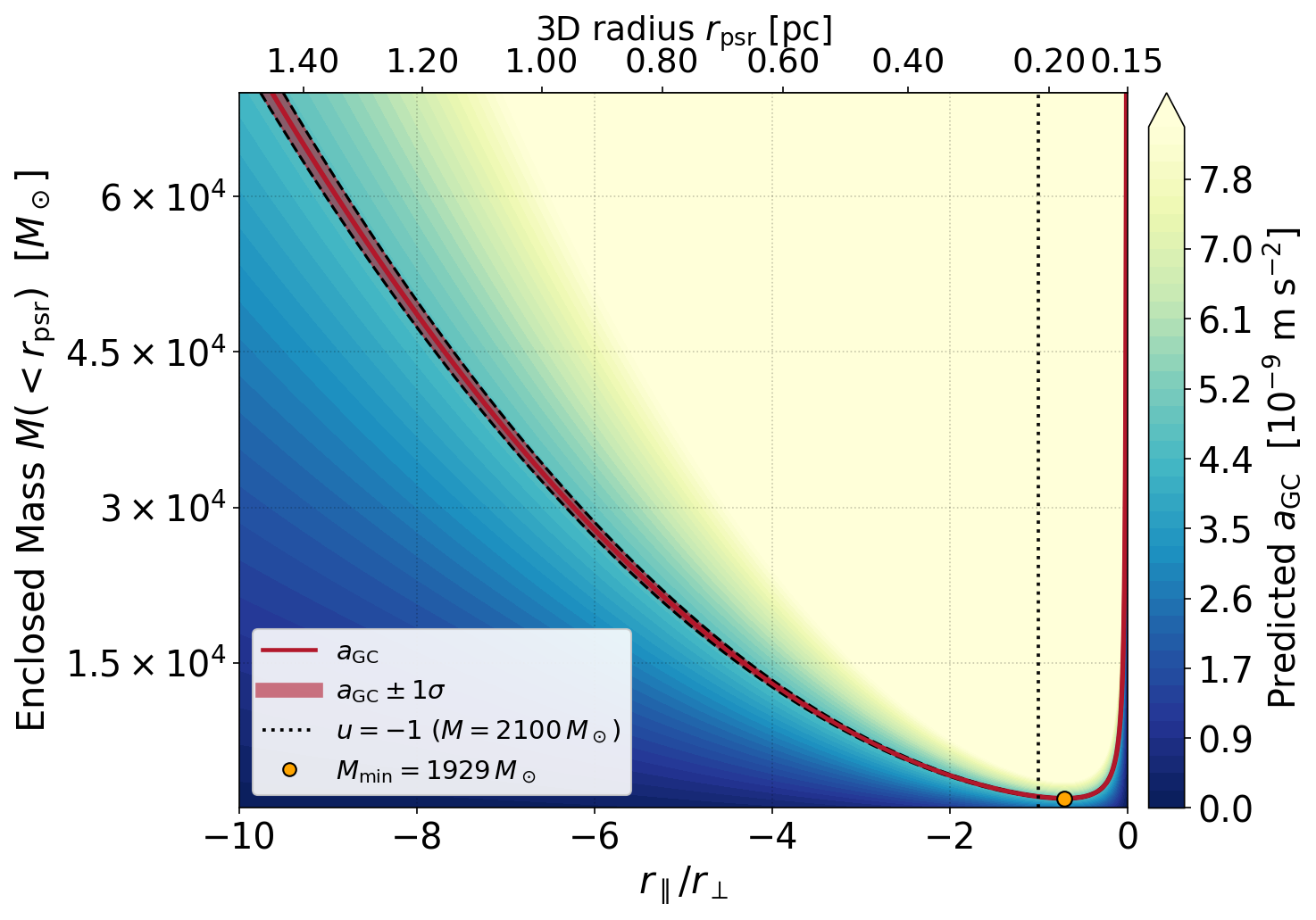} 
    \caption{Model-independent degeneracy between the enclosed cluster mass $M(r_{\rm psr})$
and the unknown line-of-sight displacement parameter $u=r_\parallel/r_\perp$
for PSR~J1807$-$2459B. Only the near side of the cluster ($u<0$) is
shown, since for the observed $a_{\rm GC}>0$ Eq.~(\ref{eq:accel}) admits solutions with $M>0$
only there. The colored background shows the predicted line-of-sight
acceleration $a_{\rm GC}$, while the red curve corresponds to the measured value
$a_{\rm GC}=(4.60\pm0.12)\times10^{-9}$\,m\,s$^{-2}$. The narrow shaded region indicates
the associated $1\sigma$ uncertainty interval. The vertical dotted black line
marks the reference configuration $u=-1$. The upper horizontal axis reports the
corresponding true 3D radial distance $r_{\rm psr}$ from the cluster center. The
orange dot marks the minimum enclosed mass. No specific dynamical or
density-profile model for the cluster is assumed.}
    \label{fig:accel_mass}
\end{figure}

Using Eq.\;(\ref{eq:accel}), we computed the full degeneracy relation between the unknown ratio $u$ and the enclosed mass required to reproduce the observed acceleration $a_{\rm GC}=(4.60\pm0.12)\times10^{-9}$ m s$^{-2}$, namely $M(r_{\rm psr})=-a_{\rm GC}\,r_\perp^2\,(1+u^2)^{3/2}/(G\,u)$. Since $a_{\rm GC}>0$, solutions with $M>0$ exist only for $u<0$: the pulsar lies on the near side of the cluster, independently of the mass profile. The resulting contour plot is shown in Fig.\;\ref{fig:accel_mass} and provides a purely geometry-driven constraint on the allowed $(u,M)$ combinations, without assuming any specific dynamical model, density slope, or velocity-dispersion profile for the cluster. The resulting lower bound on the enclosed mass is $M_{\min}=(3\sqrt{3}/2)\,a_{\rm GC}\,r_\perp^2/G=1929\,M_\odot$, obtained at $u=-1/\sqrt{2}$, with a $1\sigma$ range of $[1879,1979]\,M_\odot$ when only the uncertainty on $a_{\rm GC}$ is considered.

From the plot it is evident that a broad region of the parameter space is compatible with the observed acceleration. Moreover, since the enclosed mass cannot exceed the total mass of the cluster, $(8.15\pm0.56)\times10^4\,M_\odot$ \citep{Baumgardt_2018}, the pulsar must lie at $|u|\lesssim11$, i.e. $r_{\rm psr}\lesssim1.6$\,pc, inside the cluster half-mass radius $r_h\simeq2.24$\,pc; it therefore belongs to the inner cluster region, although it is not necessarily embedded in the very central cusp of the cluster. This model-independent construction demonstrates that the inferred acceleration does not require any fine-tuned geometry and already implies a non-negligible enclosed mass independently of the adopted cluster mass profile.

\section{B}
\label{appendix:B}

To verify independently the correlation between $\gamma$, $x$ and $\dot x$ discussed in Sec.~\ref{sec:xdot}, and its impact on the measurement of $\gamma$, we performed a Bayesian timing analysis with \texttt{TempoGravity}\footnote{\url{https://github.com/amodiocarleo/TempoGravity}}, a \textsc{Python} package that samples the parameters of the timing model with \texttt{emcee}, using \texttt{TEMPO2} to compute the residuals at each step. To speed up the analysis, we used a lighter dataset, excluding the GBT data, and a simplified noise model; for this reason, the numerical results of this appendix are not meant to be compared in detail with those of Tab.~\ref{tab:preliminary}. In this analysis, $\gamma$, $x$ and $\dot x$ were all left free, together with all the other timing parameters. 

With broad priors on all parameters, $x$ and $\gamma$ could not be constrained, as expected: when both $x$ and $\dot x$ are free, a change of $\gamma$ can be compensated by changes of these parameters, and the data alone cannot determine $\gamma$. We then imposed a narrow prior on $x$ ($\pm3\sigma$) around the value of the DDGR solution. Since in the DDGR model $\gamma$ is not a free parameter but is fixed by GR, this value of $x$ already incorporates the GR value of $\gamma$, so that a narrow prior on $x$ is in practice a narrow prior on $\gamma$ around its GR value. We indeed obtained $\gamma=14.39\pm0.06$\,ms, in perfect agreement with the GR prediction ($\simeq14.42$\,ms), with an uncertainty set mainly by the width of the prior on $x$, and strongly correlated with $x$. At the same time, $\dot x$, which absorbed the slow changes of the orbit, was consistent with zero and essentially uncorrelated with $\gamma$. These results confirm the strong correlation between $\gamma$ and $x$, and show that a precise value of $\gamma$ obtained in this way is not a measurement, but reflects the assumption made on $x$. This supports the approach adopted in Sec.~\ref{sec:xdot}, where $\gamma$ is measured with $\dot x$ fixed and the effect of the proper motion is included as a systematic uncertainty.

\begin{figure}[htbp]
    \centering
    \includegraphics[width=0.999\columnwidth]{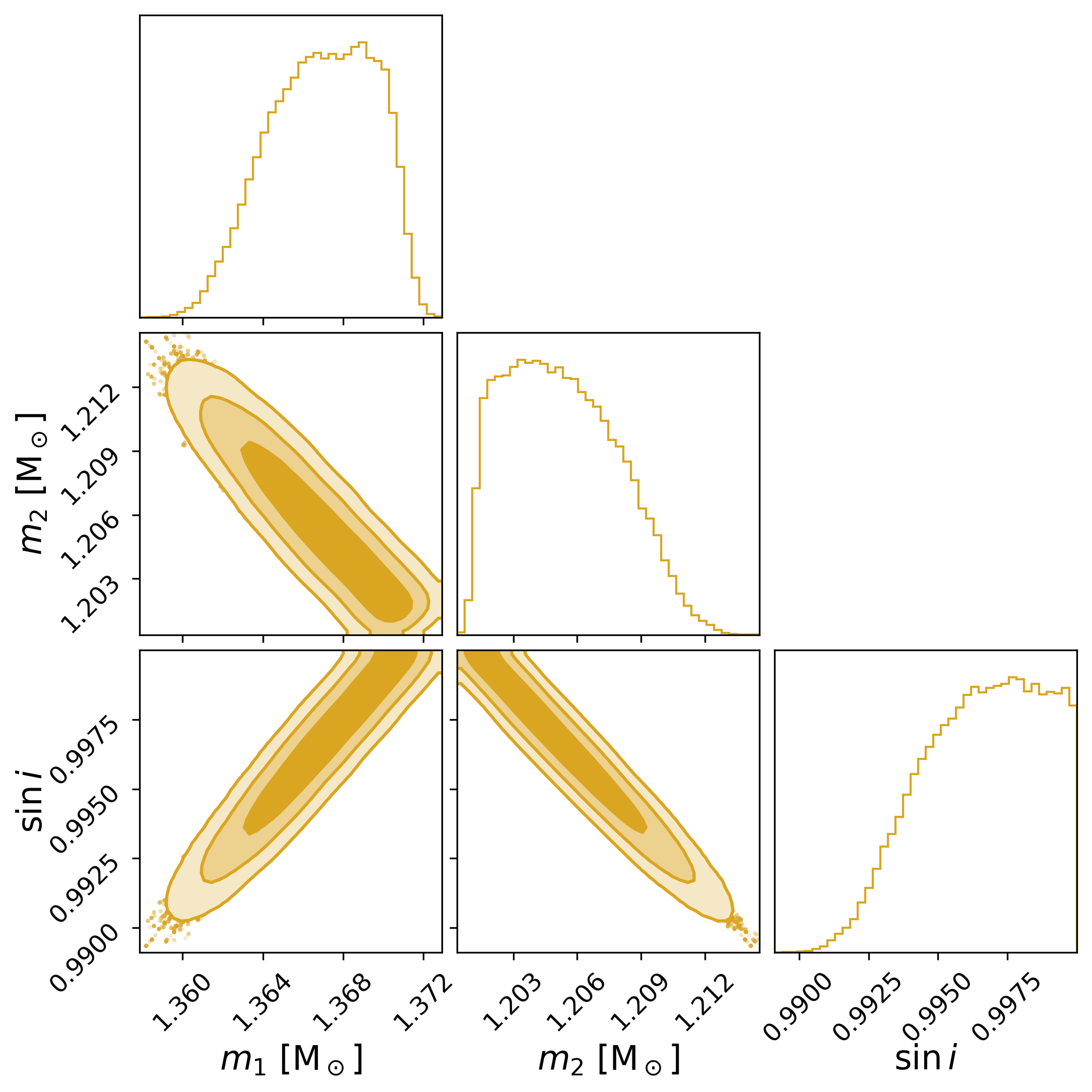}
    \caption{Posterior distributions of the pulsar mass $m_1$, the companion mass $m_2$ and $\sin i$ obtained from the \texttt{TempoGravity} analysis. The masses were derived from the posteriors of $\dot\omega$ and $\gamma$ through their GR expressions, and $\sin i$ from the mass function. Since $\gamma$ is constrained here by a prior on $x$ taken from a DDGR solution, these distributions are not independent of GR. The 2D contours represent the 68, 95 and 99.7 per cent confidence regions.}
    \label{fig:mass_corner}
\end{figure}

Combining, sample by sample, the posteriors of $\dot\omega$ and $\gamma$ through their GR expressions, we obtained mass measurements, namely $m_1=1.3671^{+0.0026}_{-0.0029}\,{\rm M_\odot}$, $m_2=1.2051^{+0.0031}_{-0.0026}\,{\rm M_\odot}$ and $\sin i=0.9968^{+0.0022}_{-0.0026}$ (Fig.~\ref{fig:mass_corner}). These values are consistent at the $1\sigma$ level with those of Sec.~\ref{sec:masses} and of \cite{lynch2012}. Therefore, once GR is assumed (here through the prior on $x$), a full Bayesian exploration of all the timing parameters, performed with a different software and on a subset of the data, yields values of $\gamma$ and of the masses that agree with the GR expectations and with the results of Sec.~\ref{sec:masses}, where only $M_{\rm tot}$, $\cos i$ and $\Omega$ were sampled. This confirms the robustness of our mass determination against the choice of the dataset, of the noise model and of the analysis method, provided that the degeneracy between $\gamma$ and $x$ is properly taken into account.

\clearpage
\onecolumngrid

\section{C}
\label{appendix:C}

\begin{figure}[!htbp]
    \centering
    \includegraphics[
        width=\textwidth,
        height=0.95\textheight,
        keepaspectratio
    ]{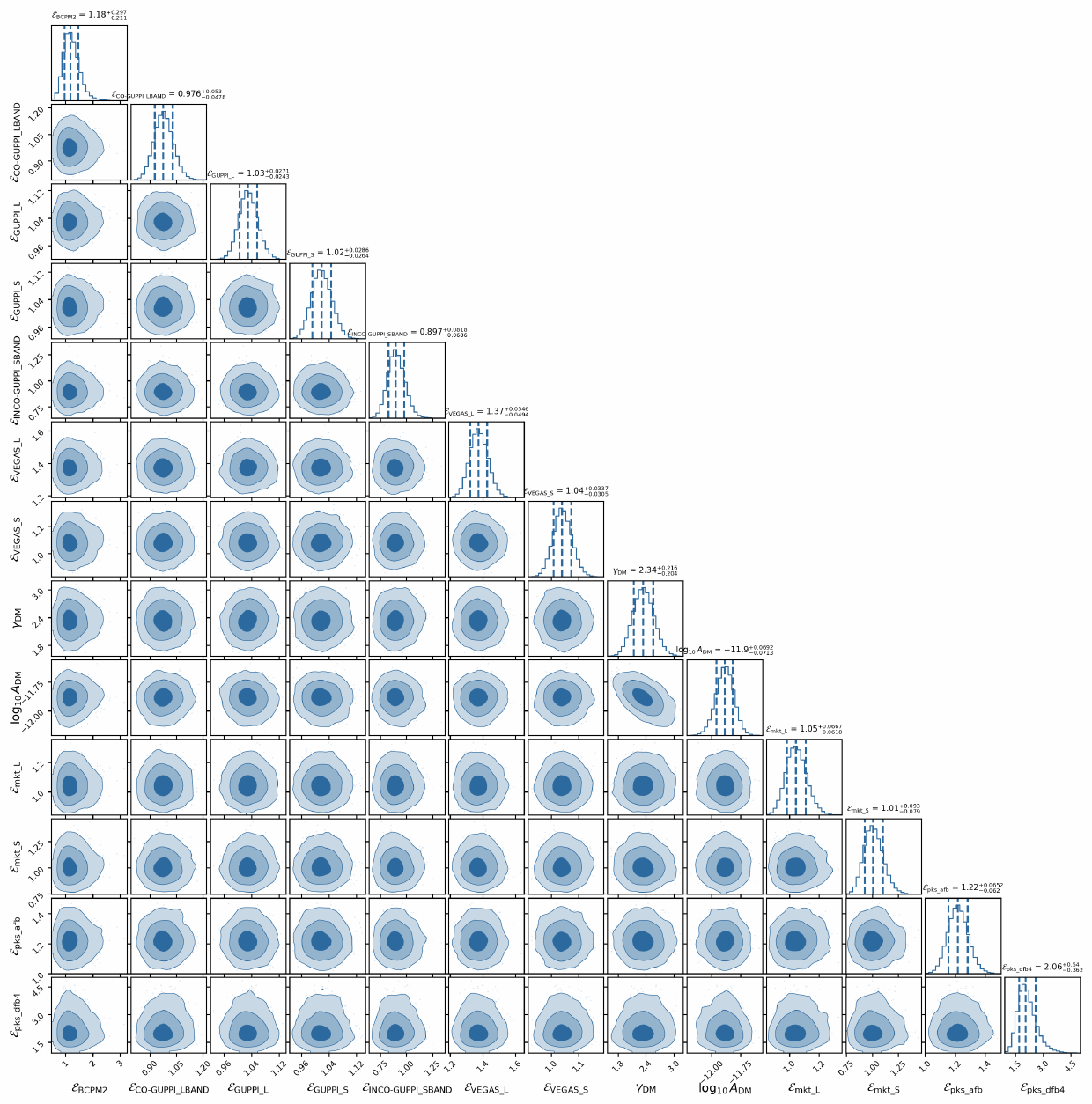}

    \caption{Corner plot showing the posterior distribution
    of the parameters of the preferred noise model.
    The diagonal elements show the marginalized 1D
    histograms for each parameter; dashed vertical lines
    mark the 16th percentile, the median (50th), and the
    84th percentile. Off-diagonal subplots show the
    correlation between the parameters and are marked by
    contours that define the 68\%, 95\% C.L.}

    \label{fig:noise_corner}
\end{figure}

\clearpage
\twocolumngrid




\end{document}